\documentclass[useAMS,usenatbib]{mnras}
\usepackage{graphicx}
\usepackage{hyperref}
\usepackage{amssymb}
\usepackage{amsmath}
\usepackage{abbrevs}
\usepackage{xspace}
\usepackage{float}
\usepackage[usenames,dvipsnames]{xcolor}
\usepackage[T1]{fontenc}
\usepackage{multicol}
\usepackage{makecell}
\begingroup
  \catcode`\_=\active
  \gdef_#1{\ensuremath{\sb{\mathrm{#1}}}}
\endgroup
\mathcode`\_=\string"8000
\catcode`\_=12

\let\oldhref\href
\renewcommand{\href}[2]{\oldhref{#1}{\hbox{#2}}}

\newcommand{\rev}{\textcolor{Black}}

\newcommand{\Msolar}{M$_{\odot}$\xspace}

\newcommand{\starsub}{$_{\normalfont\textsc{stars}}$\xspace}
\newcommand{\stars}{\textsc{stars}\xspace}
\newcommand{\turbsub}{$_{\normalfont\textsc{turb}}$\xspace}
\newcommand{\turb}{\textsc{turb}\xspace}

\newabbrev\ISM{Interstellar Medium (ISM)}[ISM]
\newabbrev\CSM{Circumstellar Medium (CSM)}[CSM]
\newabbrev\WNM{Warm Neutral Medium (WNM)}[WNM]
\newabbrev\WIM{Warm Ionised Medium (WIM)}[WIM]
\newabbrev\CNM{Cold Neutral Medium (CNM)}[CNM]
\newabbrev\IMF{Initial Mass Function (IMF)}[IMF]
\newabbrev\CMF{Core Mass Function (IMF)}[IMF]
\newabbrev\AMR{Adaptive Mesh Refinement (AMR)}[AMR]
\newabbrev\HGB{Horizontal Giant Branch (HGB)}[HGB]
\newabbrev\SFE{Star Formation Efficiency (SFE)}[SFE]
\newabbrev\TSFE{Total Star Formation Efficiency (TSFE)}[TSFE]
\newabbrev\OSFE{Observed Star Formation Efficiency (OSFE)}[OSFE]
\newabbrev\SFR{Star Formation Rate (SFR)}[SFR]
\newabbrev\YSOs{Young Stellar Objects (YSOs)}[YSOs]
\newabbrev\YSO{Young Stellar Object (YSO)}[YSO]
\newabbrev\PDF{Probability Distribution Function}[PDF]
\newabbrev\PSF{Point Spread Function}[PSF]
\newabbrev\LMC{Large Magellanic Cloud}[LMC]

\makeatletter
\newcommand*\bigcdot{\mathpalette\bigcdot@{.5}}
\newcommand*\bigcdot@[2]{\mathbin{\vcenter{\hbox{\scalebox{#2}{$\m@th#1\bullet$}}}}}
\makeatother

\makeatletter
\renewcommand\maybe@space@{%
  \maybe@ictrue % <= this is new
  \expandafter   \@tfor
    \expandafter \reserved@a
    \expandafter :%
    \expandafter =%
                 \nospacelist
                 \do \t@st@ic
  \ifmaybe@ic % <= this is new
    \space
  \fi
}
\makeatother
\title[The Indeterministic Nature of Star Formation]{On the Indeterministic Nature of Star Formation on the Cloud Scale}
\author[S. Geen]
      {Sam Geen$^{1}$\thanks{Corresponding author: Sam Geen (sam.geen@uni-heidelberg.de)}, 
      	Stuart K. Watson$^{2}$, 
      	Joakim Rosdahl$^{3}$, 
      	Rebekka Bieri$^{4}$, 
      	Ralf S. Klessen$^{1,5}$, 
      	\newauthor
      	Patrick Hennebelle$^{6}$ \\\textsl{}
{$^{1}$ Universit\"at Heidelberg, Zentrum f\"ur Astronomie, Institut f\"ur Theoretische Astrophysik, Albert-Ueberle-Str. 2, 69120 Heidelberg, Germany}\\
{$^{2}$ Department of Comparative Linguistics, University of Z\"urich, Plattenstrasse 54, CH-8032 Z\"urich, Switzerland}\\
{$^{3}$ Univ Lyon, Univ Lyon1, ENS de Lyon, CNRS, Centre de Recherche Astrophysique de Lyon UMR5574, F-69230, Saint-Genis-Laval, France}\\
{$^{4}$  Max-Planck-Institute for Astrophysics, Karl-Schwartzschild-Strasse 1, Garching, Germany}\\
{$^{5}$ Universit\"at Heidelberg, Interdisziplin\"ares Zentrum fur Wissenschaftliches Rechnen, INF 205, 69120, Heidelberg, Germany}\\
{$^{6}$ Laboratoire AIM, Paris-Saclay, CEA/IRFU/SAp - CNRS - Universit\'e Paris Diderot, 91191, Gif-sur-Yvette Cedex, France}\\}
\date{\today}

\begin{document}
\label{firstpage}
\pagerange{\pageref{firstpage}--\pageref{lastpage}}
\maketitle

% ABSTRACT 
\begin{abstract}
Molecular clouds are turbulent structures whose star formation efficiency (SFE) is strongly affected by internal stellar feedback processes. In this paper we determine how sensitive the SFE of molecular clouds is to randomised inputs in the star formation feedback loop, and to what extent relationships between emergent cloud properties and the SFE can be recovered. We introduce the \textsc{yule} suite of 26 radiative magnetohydrodynamic (RMHD) simulations of a 10,000 solar mass cloud similar to those in the solar neighbourhood. We use the same initial global properties in every simulation but vary the initial mass function (IMF) sampling and initial cloud velocity structure. The final SFE lies between 6 and 23\% when either of these parameters are changed. We use Bayesian mixed-effects models to uncover trends in the SFE. The number of photons emitted early in the cluster's life and the length of the cloud provide the strongest predictors of the SFE. The HII regions evolve following an analytic model of expansion into a roughly isothermal density field. The more efficient feedback is at evaporating the cloud, the less the star cluster is dispersed. We argue that this is because if the gas is evaporated slowly, the stars are dragged outwards towards surviving gas clumps due to the gravitational attraction between the stars and gas. While star formation and feedback efficiencies are dependent on nonlinear processes, statistical models describing cloud-scale processes can be constructed.
\end{abstract}

% ------------------------------------------------------------------------------------------------------------------------------------------------------

\section{Introduction}
\label{introduction}

Stars form in dense, turbulent, self-gravitating clouds. The most massive stars produce large quantities of energy that drive outflows from the cloud, creating time- and space-dependent feedback loops in the cloud's star formation cycle \citep[see reviews by, e.g.][]{MacLow2004,Ballesteros-Paredes2007,Klessen2016}. The path a star-forming system takes between its initial and final state depends on a number of non-linear processes that are dynamically linked. The (magneto)hydrodynamic equations and the Poisson equation governing self-gravity underlying the dynamics of the cloud are non-linear. The gravitational collapse of turbulent structures in a cloud based on these equations gives rise to a statistical distribution called the \CMF, which in turn leads to a stellar \IMF. Stars produce quantities of radiation and kinetic outflows dependent on the evolution of their atmospheres, and this feeds back into the cloud, affecting future structure evolution and star formation. Eventually, this feedback loop is terminated when all of the gas in the cloud is dispersed and only the young star cluster remains.

In this paper we randomise the sampling of the stellar \IMF and the turbulent velocity field of a single star-forming cloud, determine how much this affects the final mass in stars formed by the cloud, and identify whether there are any emergent linear trends between the early state of the cloud and the final mass of stars formed.

A number of previous works exist to explain the expansion of photoionised HII regions in molecular clouds using analytic arguments \citep{KahnF.D.1954,SpitzerLyman1978,Whitworth1979,Franco1990,Williams1997,Hosokawa2006,Raga2012,Geen2015b}. These have been applied to explaining the structure of observed HII regions in, e.g., \cite{Didelon2015} and \cite{Tremblin2014a}. Analytic models that describe the regulation of star formation by stellar feedback also exist \citep{Matzner2002,Krumholz2006,Goldbaum2011}. \rev{The problem has also been studied using 3D numerical simulations \citep[among others]{Dale2005,Gritschneder2009,Peters2010,Walch2012,Dale2012,Colin2013,Howard2016,Geen2017,Howard2018,Ali2018}, with varying parameter spaces and physical models being studied}. \cite{Colin2013} argues that local compactness alters the resulting \SFE, while \cite{Dale2012} find systematic trends in the global cloud properties and \SFE.

Coupling the 1D analytic and 3D hydrodynamic approaches into a single theory of star formation is somewhat difficult. Analytic models must make certain simplifications that omit features of the full 3D approach, with the ansatz that these features are unimportant in describing the full cycle of star formation in clouds. Meanwhile, in numerical simulations it is much harder to understand the underlying causal relationships between the modelled system components due to the complexity of the full set of equations being solved. Simulations are considerably more expensive than simpler models, and performing multiple simulations that capture the full parameter space of the problem is often prohibitively expensive.

The goal of this paper is to explore how sensitive star formation is to small changes in model inputs by performing multiple realisations of the same cloud with different choices of randomly sampled input parameters, and identifying relationships between input and emergent properties of the cloud and the final \SFE. 

The problem of statistical variation in astrophysical systems has been studied in different domains. \cite{Goodwin2004a} investigate the effect of levels of turbulence on the formation of N-body star cluster simulations. Meanwhile, \cite{Girichidis2011}, \cite{Girichidis2012a} and \cite{Girichidis2012b} explore the role of cloud properties such as the density profile and turbulence inside the cloud on resulting cloud properties. The \textsc{slug} framework \citep{DaSilva2012,DaSilva2014,Krumholz2015} has quantified the effect of stochastic IMF sampling on cluster properties such as \SFR tracers and photometry. \cite{Hoffmann2017} investigate the problem in the context of planet-forming disks. On a larger scale, \cite{Keller2018} study the effect of stochastic noise on cosmological galaxy formation simulations.

We focus in this case on the mass of massive stars sampled from the stellar \IMF and the seeding of the initial turbulent velocity field in the cloud. We ask two broad questions. 

\textit{One, for a given system, how much variation in the resulting global properties of the cloud is there when these two processes are sampled differently?}

\textit{Two, can we na\"ively recover any trends from the resulting cloud structure, or is the evolution of molecular clouds dominated by nonlinear processes?}

We split this paper into the following sections. In Section \ref{simulations}, we present the simulations performed in this study and the numerical setup used. In Section \ref{globalprops}, we discuss the global properties of each of our simulations and the evolution of the clouds over time. In Section \ref{correlations} we identify trends in the cloud using a statistical model and suggest which emergent properties of the system can be used to explain the resulting \SFE of the cloud. Finally, in Section \ref{discussion}, we discuss our results and provide scope for future work on this topic.

\section{Numerical Simulations}
\label{simulations}

\begin{table*}
\begin{center}
\begin{tabular}{llllllll}
\textbf{Clouds with Varying IMF Sampling (Subscript ``\stars'')} \\
\hline
Stekkjarstaur (STE\starsub), Giljagaur (GIL\starsub), St\'ufur (STU\starsub), \TH{}v\"orusleikir (THV\starsub), Pottaskefill (POT\starsub), \\ Askasleikir (ASK\starsub), Hur\dh{}askellir (HUR\starsub), Skyrg\'amur (SKY\starsub), Bj\'ugnakr\ae{}kir (BJU\starsub), \\ Gluggag\ae{}gir, (GLU\starsub), G\'atta\th{}efur (GAT\starsub), Ketkr\'okur (KET\starsub), Kertasn\'ikir (KER\starsub) \\
\hline
\textbf{Clouds with Randomly Generated Initially Turbulent Velocity Fields (Subscript ``\turb'')} \\
\hline
Gr\'yla (GRY\turbsub), J\'olak\"otturinn (JOL\turbsub), Joulupukki (JOU\turbsub), G\"avlebocken (GAL\turbsub), Ti\'o de Nadal (TIO\turbsub), \\ Caganer (CAG\turbsub), Snegurochka (SNE\turbsub), Befana (BEF\turbsub), St Lucy (STL\turbsub), Mari Lwyd (MAR\turbsub), \\ Tante Arie (TAN\turbsub),Old Man Bayka (OLD\turbsub), Y\'agena Ab\~at (YAG\turbsub) \\

\end{tabular}
\end{center}
\caption{List of simulation names used in this paper. The names have no preferential ordering since the quantities being varied are sampled at random. Shortened names used for referencing are given in parentheses. The \protect\stars simulations are run using the same turbulent initial velocity fields as in \protect\cite{Geen2017}. The \turb simulations use the same IMF sampling as in the HUR\protect\starsub simulation. Lists of stellar masses appearing in each cloud in the \protect\stars set are given in Appendix \protect\ref{appendix:starlist}.}
\label{simtable} 
\end{table*}

In this Section we introduce the \textsc{yule} suite of simulations used in this paper. These are listed in Table \ref{simtable}. The suite is divided into two groups of 13 simulations, labelled \stars and \turb\footnote{The \stars simulations are named after the Icelandic \textit{j\'olasveinarnir}, while the \turb simulations are named after other winter figures.}. All of these simulations are given names with three letter codes, since there is no preferential ordering to the randomly sampled simulation parameters. For all of the simulations in this paper we use the radiative magnetohydrodynamic Eulerian \AMR code \textsc{RAMSES} \citep{Teyssier2002,Fromang2006,Rosdahl2013}.

\subsection{Initial Conditions}

Each of our simulations has an identical initial setup. Each cloud has an initial gas mass of $10^4$ \Msolar. The density distribution is identical to the ``L'' cloud in \cite{Geen2017}. In that paper, we show that the cloud's density structure is very similar to that of clouds in the nearby Gould Belt \citep{Poppel1997a}, $<$ 500 pc away from the Sun. The cloud has an initially spherically symmetric density profile, with an isothermal profile up to a radius $r_{ini}=7.65$ pc. See \cite{Iffrig2015}, \cite{Geen2015b} and \cite{Geen2016} for more detailed information about the initial density profile. A uniform sphere is placed out to $2~r_{ini}$ with a density 0.1 times that just inside $r_{ini}$. The total length of the cubic volume simulated is 16 times $r_{ini}$, or 122 pc.

The cloud is initially virialised, with a turbulent velocity field. The initial field has a Kolmogorov power spectrum ($P(k) \propto k^{-5/3}$) with random phases. We generate 13 velocity fields, each with a different random seed. Labels identifying these fields are listed in Table \ref{simtable}. The global freefall time in our cloud $t_{ff}$ is 4.2 Myr. 

The initial ratio of $t_{ff}$ to sound crossing time $t_{sound}$ on the cloud scale is 0.15 (note that for Jeans collapse, $t_{ff} < t_{sound}$). The ratio of $t_{ff}$ to the crossing time of turbulent flows $t_{RMS}$, defined by the RMS velocity $V_{RMS}$, is 2. 

The initial ratio of $t_{ff}$ to Alfv\'en crossing time in the cloud is 0.2. The Alv\'en crossing time is defined as $R \sqrt{4 \pi \rho} / B$, so a higher ratio for a given cloud means a larger magnetic field. \rev{The precise value of $B$ is density dependent and evolves as gravity and HII regions compress structures in the cloud. Initially, the magnetic field is pointed along one coordinate axis with a peak strength of 457 $\mu$G, though over time the peak field strength rises to around 100 mG in the densest structures \citep[][find magnetic field strengths of a few hundred mG around massive stars in Orion]{Pellegrini2009}. We will explore the evolution of the magnetic field in the simulation more fully in future work.}

\subsection{Cloud Evolution and Sink Formation}

Each simulation is performed on an adaptively refined octree, in which each cell in a cubic grid with size $2^{l_{min}}$ is recursively sub-divided into 8 evenly-sized cells up to an effective resolution of $2^{l_{max}}$ cells. Here, the minimum level $l_{min}=7$, giving a root grid of $128^3$ cells, mainly designed to capture a large empty volume around the cloud that traces outflows. We refine a sphere of diameter $8 r_{ini}$ for a further 2 levels. For an additional 3 levels, up to $l_{max}=12$, any cell in the simulation volume is refined if it is either Jeans unstable by a factor of 10 over the Jeans stability limit, or more massive than 0.25 \Msolar. The maximum spatial resolution is 0.03 pc, or $122~\mathrm{pc} / 2^{12}$. At $t_{ff}$, our simulations typically have $\sim10^6$ cells at level 9 ($\Delta x=0.24~$pc) and $\sim20000$ cells at the highest level of refinement.

The cloud is ``relaxed'' by simulating without self-gravity for 0.5 $t_{ff}$, so that the turbulent velocity and initially spherically symmetric density fields can couple \citep[see][amongst others]{Klessen2000,Lee2016a}. After this time we apply self-gravity to the cloud.

Once gas cells reach at least 0.1 of the Jeans density at the highest refinement level, we identify them in clumps. This is done by identifying contours from high to low density via the ``watershed'' method, and then merging clumps identified in this first pass into larger structures by locating saddle points in the density field \citep{Bleuler2014}. If a clump exceeds the Jeans density, it forms a sink particle, which accretes 90\% of the mass above this density threshold \citep{Bleuler2014a}. \rev{Sink particles have a variable mass since they accrete over time, but the average sink mass by the end of a simulation is between 50 and 100 \Msolar.}
 
\subsection{Radiative Transfer and Cooling}

Ionising photons are propagated across the full adaptive grid using the M1 method described in \cite{Rosdahl2013}. We use three `grey' photon groups to bin the full spectrum of ionising photons. These bins have lower bounds at the ionisation energies of HI, HeI and HeII. Each AMR grid cell tracks the photon density and flux of photons in each group. We couple the photons to the gas at every timestep and follow the ionisation states of hydrogen and helium in every AMR grid cell.  \textsc{Ramses} requires a single photon energy and cross-section (weighted by photon energy and number) for each photon bin, so we pick representative values for each of them based on a stellar population, using identical values to \cite{Geen2017}. 

The number of photons we inject around each source is described in Section \ref{simulations:stellarevolution}. We do not include photon energies lower than the ionisation energy of hydrogen, or direct or scattered radiation pressure. We discuss these omissions in Section \ref{discussion:limits}.

We use the radiative cooling module described in \cite{Geen2016,Geen2017}. For gas in collisional ionisation equilibrium, we use \cite{Audit2005} below $10^4~$K and a fit to \cite{Sutherland1993} above this limit. We perform cooling on photoionised hydrogen and helium as described in \cite{Rosdahl2013}, with an additional fit to \cite{Ferland2003} to capture photoionised metal cooling. All cells are assumed to be at solar metallicity.

\subsection{Massive Star Formation}

In \cite{Geen2017}, we used a fit to a population model for UV photon emission to limit issues with stochasticity in our results. In this paper we instead sample individual massive stars to study the role of statistical variation in the sampling of the IMF and to track individual massive stars within the cluster. 

In our simulations we track the mass distribution of all stars via sink particles. In addition, we follow the evolutionary state of stars above 8 \Msolar in order to track how much ionising radiation the stars produce. Below this mass stars do not produce significant quantities of ionising radiation. The initial masses of the stars are distributed according to a Chabrier \IMF \citep{Chabrier2003}. In order to provide reproducible results, we pre-generate 13 lists of stellar masses by randomly sampling from the \IMF until we reach $10^4$ \Msolar. We extract all stars above 8 \Msolar in the order they were first sampled. Stars below this mass are accounted for as mass in sink particles. Every time the code generates a massive star, it draws from this list in order and is synchronised such that the same star is not selected by different CPUs in the same timestep. The lists of stars formed in each simulation are given in Appendix \ref{appendix:starlist}.

Each sink particle tracks the amount of mass accreted onto it with the variable $\Delta m_{sink}$, where the sum of $\Delta m_{sink}$ over all sink particles is $\Delta m_{*}$. Every time $\Delta m_{*}$ exceeds 120 \Msolar, we identify the sink particle with the largest $\Delta m_{sink}$. We create a ``virtual'' stellar object, representing a single massive star. This stellar object is attached to the sink particle and is position is always set to be the same as the sink's. The sink's $\Delta m_{sink}$ is then decremented by 120 \Msolar and the process is repeated. We decrement 120 \Msolar rather than the mass of the stellar object in order to account for stars below 8 \Msolar in the mass distribution of the sinks. We use steps of 120 \Msolar since in the \IMF used there is one star above 8 \Msolar per 120 \Msolar of stars formed.

This method for sampling massive stars has been previously used by, e.g., \cite{Hennebelle2014}, \cite{Gatto2017} and \cite{Peters2017}, although these authors create new stars when sinks accrete 120 \Msolar on a per-sink basis rather than in the cluster as a whole, as we do. The reason for this is that they simulate a kpc-wide section of a galactic disk, which has more mass and lower spatial resolution than our cloud-scale simulations.

We have compared this method to, e.g., the Poisson sampling described in \cite{Sormani2016}, and find properties such as the ionising photon emission rate converge for large cluster masses. We discuss the consequences of this choice of massive star generation model in Section \ref{discussion:limits}. 

\subsection{Stellar Evolution and Feedback Model}
\label{simulations:stellarevolution}

We track the age of each massive star in the stellar objects described above, and the position of the sink it is attached to. At each timestep we deposit photons in each energy bin described above onto the grid at the position of the sink and allow them to propagate. The emission rate of photons from a star of mass $M_s$ is calculated using a time-dependent stellar evolution model. 

We extract spectra for individual stars from \textsc{Starburst99}, \cite{Leitherer2014} using the rotating tracks from the Geneva model, \cite{Ekstrom2012} at solar metallicity. This is similar to the approach taken by \cite{DaSilva2012} with more up-to-date stellar evolution tracks. Sample values for this model are given in Appendix \ref{appendix:starmodel}. 

We integrate over the spectra in each photon energy bin to create a set of evolutionary tracks for each star at an interval of 5 \Msolar, from 5 to 120 \Msolar. In order to interpolate correctly between stellar tracks with different lifetimes, we divide the time in each track by the total lifetime of the star and interpolate between normalised stellar ages to find the photon emission rate for the star.

All of the simulations are stopped at the point that the first star in the cluster reaches the end of its lifetime (and should go supernova), or the point at which the mass in sink particles (here taken to be the cluster mass) plateaus, whichever happens earlier. We discuss the predicted contributions from the missing feedback processes to the cloud's evolution in Section \ref{discussion:limits}.

We do not, in this first instance, implement other feedback processes such as stellar winds or supernovae, as in, e.g. \cite{Dale2014}, \cite{Rey-Raposo2016}, \cite{Gatto2017} or \cite{Peters2017}. These processes heat the gas to high temperatures ($\sim 10^8 - 10^9~$K) and significantly increase the cost of the simulations over simply including photoionisation, which heats the gas to an equilibrium temperature of $\sim10^4~$K.

\section{Global Cloud Properties}
\label{globalprops} 

In this section we present an overview of the behaviour of each of the clouds in the \textsc{yule} suite of simulations and their physical properties.

\subsection{SFE versus IMF Sampling}
\label{imfsampling}

\begin{figure*}
	\centerline{\includegraphics[width=0.94\hsize]{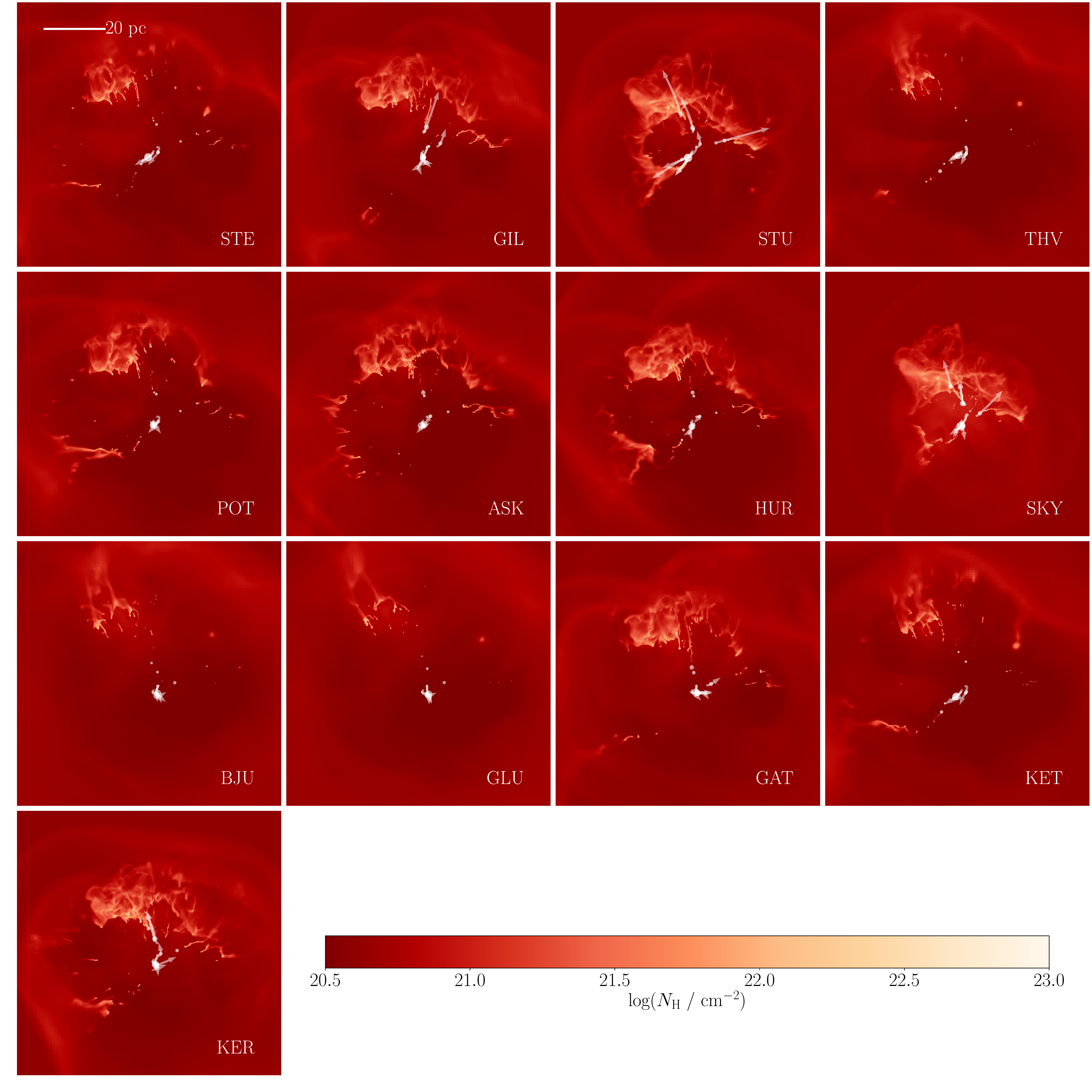}}
	\caption{Hydrogen column number density $N_H$ maps of each simulation in the \protect\stars set of simulations, where the IMF sampling is varied but the initial cloud structure is the same. Each image is projected along the $z$ axis at $t=1.5t_{ff}$, i.e. $t_{ff}$ after gravity is first turned on in the simulation. Each image is 78 pc across, zoomed into the central 70\% of the full box size (112 pc). White dots show the location of sink particles, with dot size proportional to sink mass. White lines show the track followed by each sink particle containing a massive star during the star's lifetime, with arrows at the end of each track.}
	\label{imfvis}
\end{figure*}

In Figure \ref{imfvis} we plot the gas column density and sink particle positions for the 13 \stars simulations (in which we vary the \IMF sampling used) at time 1.5 $t_{ff}$, i.e. $t_{ff}$ after gravity is turned on, where $t_{ff}=4.2$ Myr is the freefall time for the cloud as a whole. In all cases there is a well-developed HII region. The white tracks represent where sinks containing stellar objects (i.e. our sampled massive stars) travel over the course of their lifetime. See Section \ref{correlations:physical} for discussion of which simulations have stars that travel furthest. 

\subsection{SFE versus Initial Velocity Field}
\label{initialvelocity}

\begin{figure*}
	\centerline{\includegraphics[width=0.94\hsize]{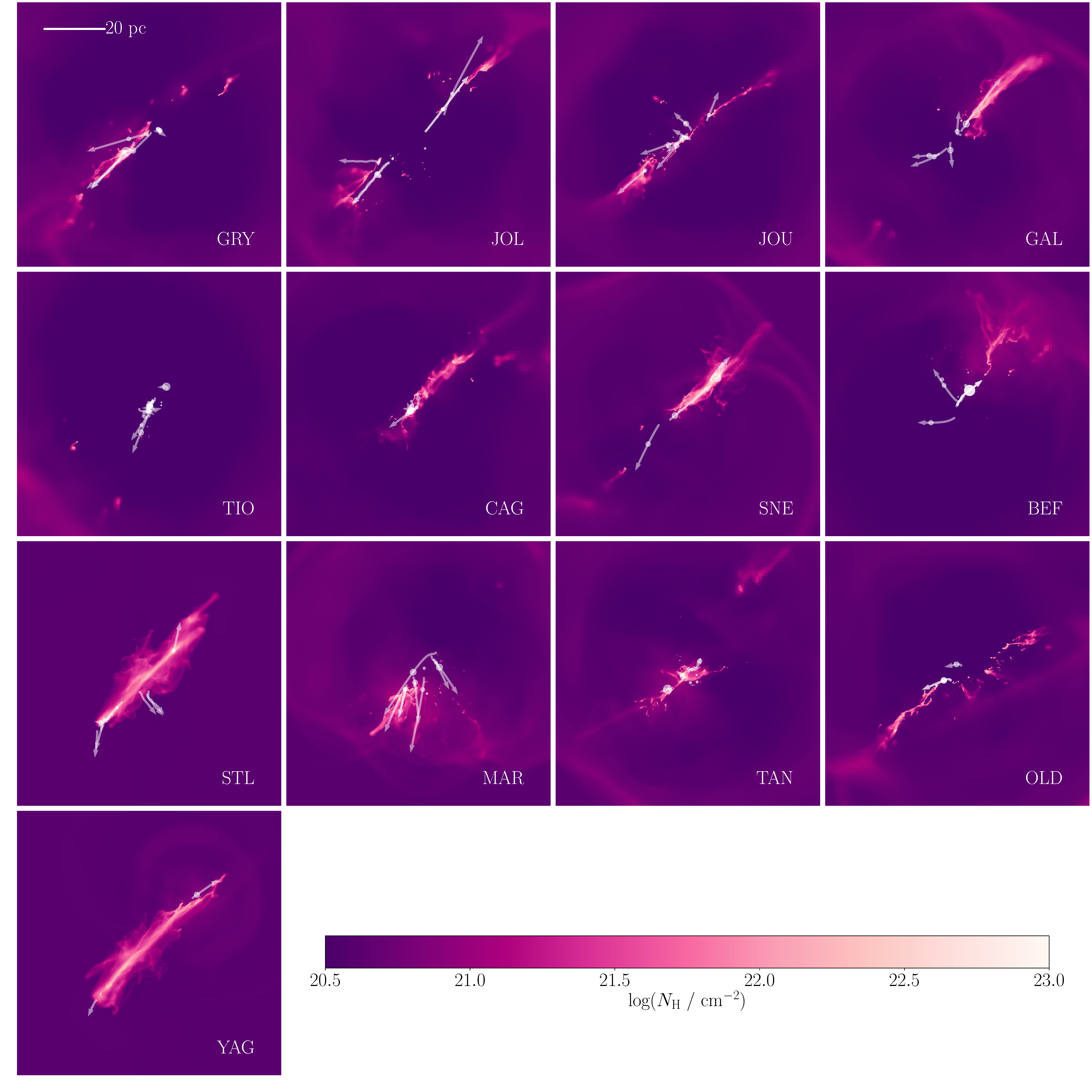}}
	\caption{Column density $N_H$ maps of each simulation in the \protect\turb set of simulations, where the IMF sampling is kept the same but the initial turbulent velocity field is generated with a different random seed. The images are constructed as in Figure \ref{imfvis}.}
	\label{icvis}
\end{figure*}

In the \turb set of simulations we use the same \IMF sampling but vary the random seed used to generate the initial turbulent velocity field. In Figure \ref{icvis} we plot all of these simulations $t_{ff}$ after the gravity is first turned on in the simulations. All of the clouds display a directionality, which is governed by the largest mode of the power spectrum used to generate this field. However, differences in the smaller modes change the small scale structure of the cloud (see Figure \ref{tsfe}), and where the stars form, with some clusters being centrally concentrated and others being spread out along the longest axis of the cloud, as well as the resulting final \SFE. Simulations YAG\turbsub and STL\turbsub show relatively little star formation after $t_{ff}$, while in TIO\turbsub nearly all of the dense gas has been dispersed by ionising radiation from the cluster.

\subsection{Total Star Formation Efficiency}
\label{globalprops:sfe}

\begin{figure*}
	\centerline{\includegraphics[width=0.96\hsize]{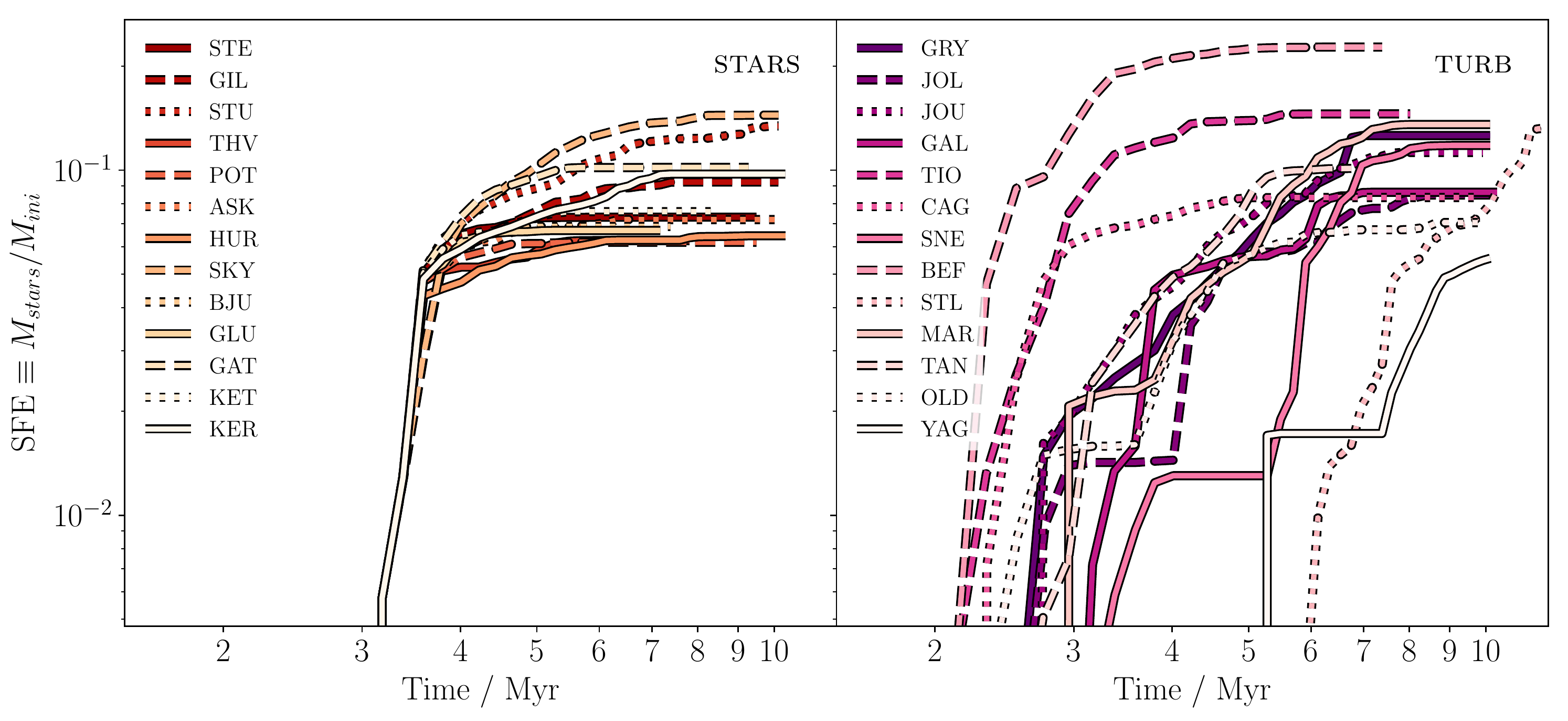}}
	\caption{Total Star Formation Efficiency (SFE) versus time for each simulation, given by the total mass in stars (sink particles) divided by the initial cloud mass. As in Figure \protect\ref{nphotons}, on the left are simulations in which we vary the mass of massive stars formed. On the right are simulations where the initial random seed of the turbulent velocity field is varied.}
	\label{tsfe}
\end{figure*}

In Figure \ref{tsfe} we plot the Total \SFE of each simulation over time, which is calculated as the fraction of total initial gas mass of the cloud ($10^4$ \Msolar) that has been converted into sink particles. The sink particles represent the young star cluster. In the \stars simulations, the final \SFE varies by a factor of $\sim$ 3 from 6 to 15\%, while in the \turb simulations the final \SFE varies from 5 to 23\%, though one of the lower \SFE simulations is still forming stars at the point at which we stop the simulation. The time at which star formation begin varies in the \turb simulations, an effect also found in \cite{Girichidis2012a}. The \stars simulations are identical before feedback begins, and so there is no variation in the time the first star is formed.

In the \turb runs, there is considerable difference between the times at which stars form and the star formation rate. In the \stars runs, since the initial structure of the cloud is the same and so the only difference is the number of photons emitted by the stars sampled from the \IMF. The choice of stellar masses in the cluster produces less scatter in the \SFE than the initial distribution of turbulent structures in the cloud. \cite{Girichidis2012a} also find that the initial gas structure is highly important in setting the star formation rates of molecular clouds. 

In the \stars simulations, the median \SFE is 7.3\% $\pm$ 1.5\%. In the \turb simulations, the median \SFE is 11.3\% $\pm$ 2.4\% (the errors here are the interquartile range, IQR). Note that these values are for individual studies changing a single randomly sampled effect, and the actual spread in \SFE is likely to be larger as effects (IMF and turbulent velocity sampling) are combined. There are also some outliers, particulary BEF\turbsub, which has a \SFE 50\% higher than the next highest (TIO\turbsub) and 100\% higher than the median. It is possible that with more than 13 simulations in each set we would find much higher and lower values.

\rev{In \cite{Geen2017}, where we simulate an identical cloud to the \stars set with and without feedback, the cloud without feedback tended towards a 100\% star formation efficiency over time. As authors such as \cite{Federrath2014} have shown, turbulence, magnetic fields and jets can suppress star formation rates, however in an isolated system with no external forces, only feedback can prevent a SFE close to 100\% as an end state.}

These results suggests that, na\"ively, the \SFE of a cloud similar to those in the nearby Milky Way environment \citep{Geen2017} is difficult to predict with reasonable accuracy. In Section \ref{correlations}, we determine whether linear relationships between initial cloud properties and the final \SFE can be recovered.

\subsection{Ionising Photon Emission Rate}
\label{globalprops:nphotons}

\begin{figure*}
	\centerline{\includegraphics[width=0.96\hsize]{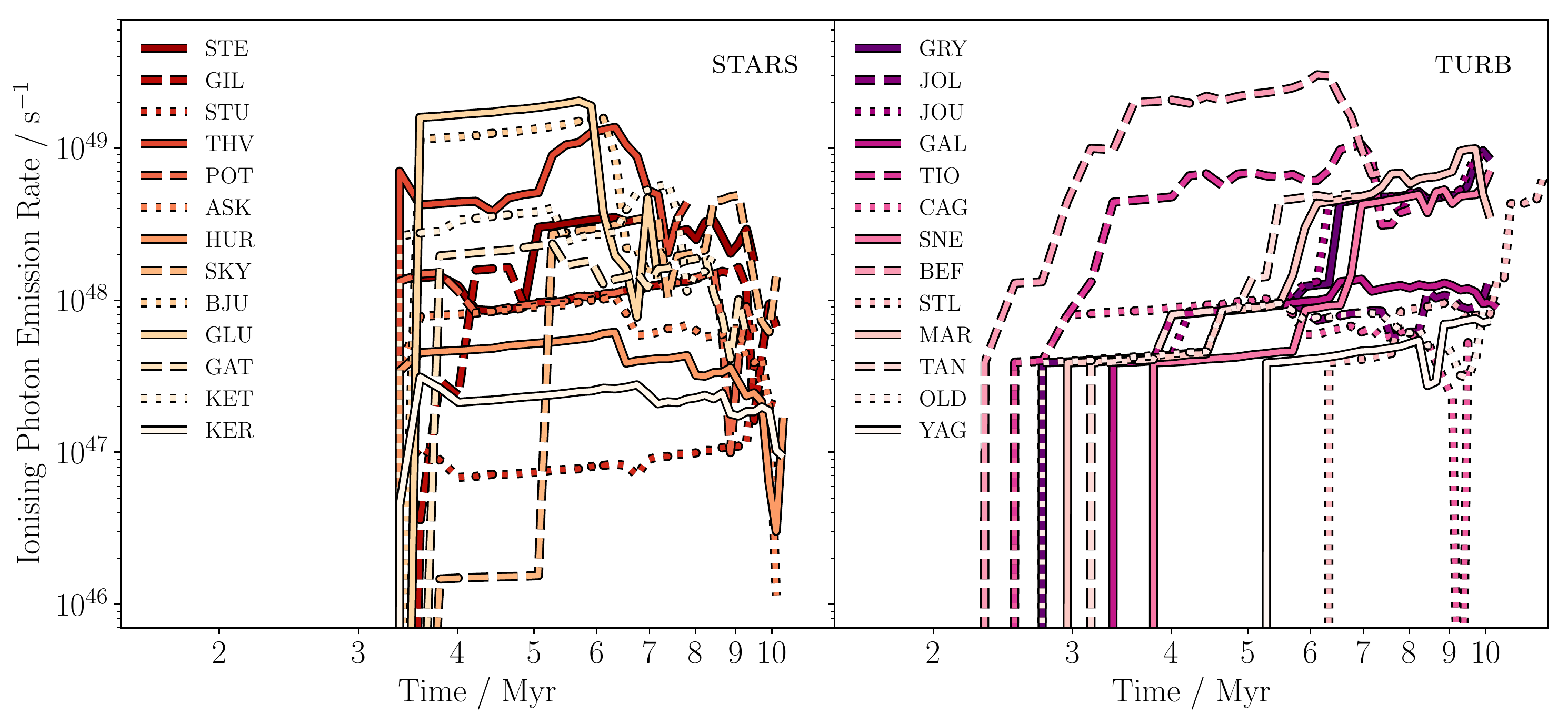}}
	\caption{Number of ionising photons, summing over all radiation bins, emitted per unit time by the cluster as a whole versus time for each simulation. On the left are simulations in which we vary the mass of massive stars formed. On the right are simulations where the initial random seed of the turbulent velocity field is varied. We stop the simulations at the point when the first star dies, since we do not explicitly include a supernova model in these simulations.}
	\label{nphotons}
\end{figure*}

We plot the total emission rate of ionising photons in Figure \ref{nphotons}. Our simulations implement a time-dependent ionising photon emission rate based on stellar evolution models (see Section \ref{simulations:stellarevolution}). As such, the number of photons emitted by a star will change over time. The global ionising photon emission rate will also increase as new stars are formed. This can be seen in the \turb simulations shown in the right panel of Figure \ref{nphotons}, where we use the same stellar masses whenever a new massive star is formed. A larger \SFE thus results in more photons being emitted. The total ionising photon emission rate is variable over time due to the stellar evolution model used (see Appendix \ref{appendix:starmodel}), which complicates the picture. We discuss the link between the number of photons emitted and the \SFE in Section \ref{correlations}.

\subsection{HII Region Radius}
\label{globalprops:radius}

\begin{figure*}
	\centerline{\includegraphics[width=0.96\hsize]{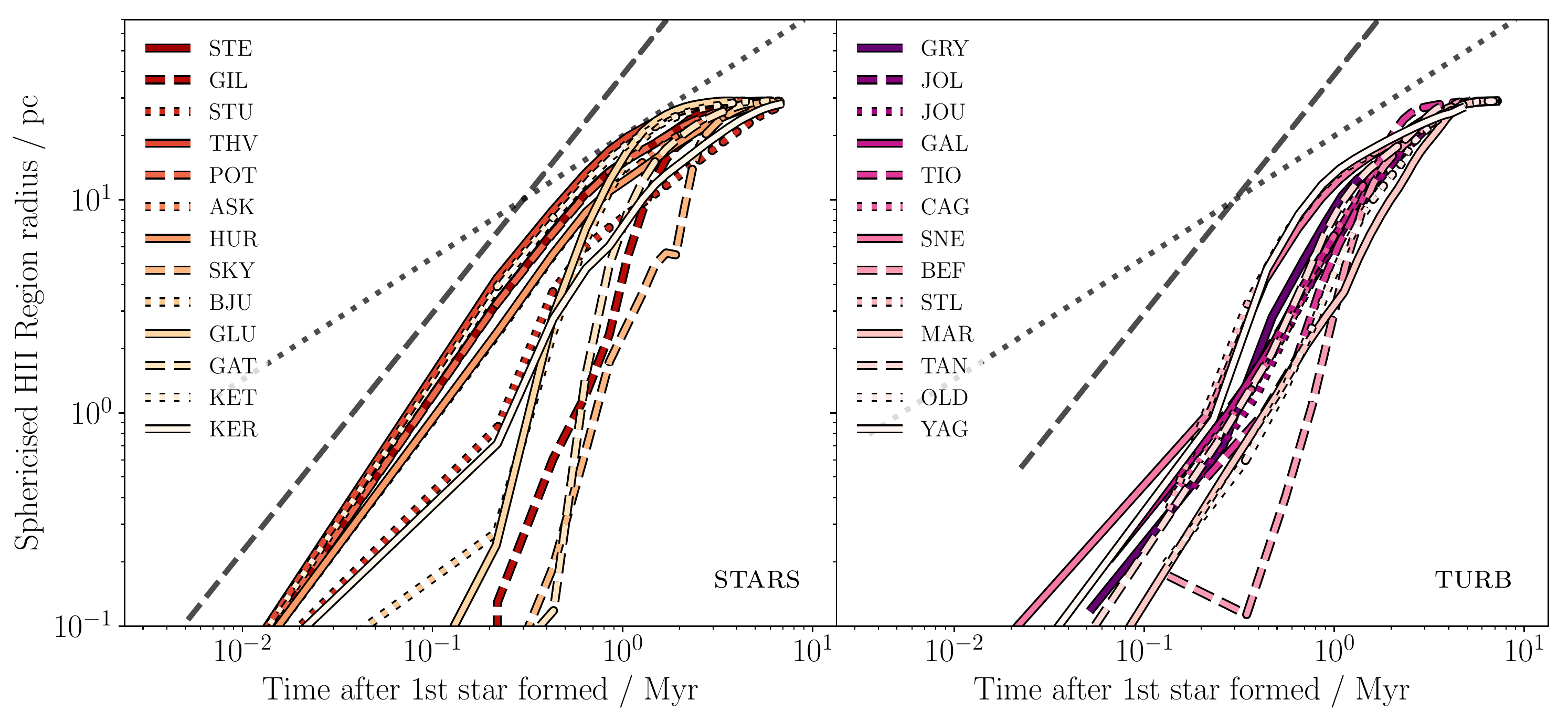}}
	\caption{Radius of the HII region(s) in the cloud as a function of time. The radius is calculated as follows. We remove all cells with an ionisation fraction $x_{HII}$ lower than 0.1. We then calculate the volume of the ionised gas in these cells as $V = \sum_{i} V_i x_{HII,i}$, where $V_i$ is the volume of the cell $i$ and $x_{HII,i}$ is the ionisation fraction in the cell. Finally, we calculate the radius as $(3 V / 4 \pi)^{1/3}$. As in Figure \ref{nphotons}, on the left are simulations in which we vary the IMF sampling. On the right are simulations where the initial random seed of the turbulent velocity field is varied. Note that above 60 pc (half the box length), flows begin to leave the box. Radii higher than this are diagonal to the Cartesian axes. We overplot the gradient of Equation \protect\ref{powerlawradius} (i.e. the normalisation is arbitrary, since it depends on various other factors). We solve Equation \protect\ref{powerlawradius} for a flat density profile (dotted black line) and for a profile of $w=1.71$, the power law fit to Figure \protect\ref{rayprofs} (dashed black line). We do this to illustrate the time dependence of the solution.}
	\label{radiusinsnap}
\end{figure*}

In Figure \ref{radiusinsnap}, we plot the radius of the HII region. We compare this to the time-dependence of the analytically-derived expansion rate given in Appendix \ref{appendix:ionfront}. In order to find a simple, robust HII region radius measurement, we locate all cells with hydrogen ionisation fraction $x_{HII} > 0.1$. We then find their ionised volume $V = \sum_{i} V_i x_{HII,i}$, where $V_i$ is the volume of the cell $i$. We then calculate the radius as $(3 V / 4 \pi)^{1/3}$.

In the same Figure, we overplot the slopes of the analytic model in Equation \ref{powerlawradius} assuming a flat (dotted line) and power law (dashed line, with index -1.71) density field as fitted to Figure \ref{rayprofs}. We set $t=0$ as the time the first star is formed. Note that the expansion close to the edge of the box ($\simeq$60 pc) and beyond is inaccurate because material begins to leave the simulation volume (larger radii are found along directions diagonal to the Cartesian axes).

HII regions in denser clouds can stall or collapse  due to pressure from gravoturbulence \citep[see][for models of this process]{Geen2015b,Geen2016}. This stalling is less important in the clouds modelled in this study due to the relatively low density of the simulated clouds \citep[see][]{Geen2017}. For much denser clouds, we expect the ambient cloud medium to resist the HII region expansion more strongly.

The power law solution for the radial expansion (dashed line) is a good match to the simulated radial expansion rate. The uniform cloud solution (dotted line) is shallower than the early expansion, suggesting that indeed the expansion of HII regions should be described by considering the power law density profile around the star(s) rather than measuring the average density of the cloud.

Note that we simply overplot the time dependence of the power law solution in Equation \ref{powerlawradius}, rather than a full solution for all simulations given in \cite{Geen2015a}. There are some variations since the emission rate of photons changes (see Figure \ref{nphotons}), introducing an extra time dependence. 

The ionisation front reaches the edge of the cloud at 10 pc after roughly 0.3 to 2 Myr. However, dense clumps remain embedded and can continue to accrete. The position of the clumps in the cloud in relation to the position of ionising radiation sources is thus key to understanding the final \SFE of the cloud. We introduce models for this in \cite{Geen2015a}, based on the clump evaporation models of \cite{Bertoldi1990}, though a more accurate model requires a more detailed understanding of where the clumps are as a function of time with respect to the ionising sources.

\subsection{Total Momentum}
\label{globalprops:momentum}

\begin{figure*}
	\centerline{\includegraphics[width=0.96\hsize]{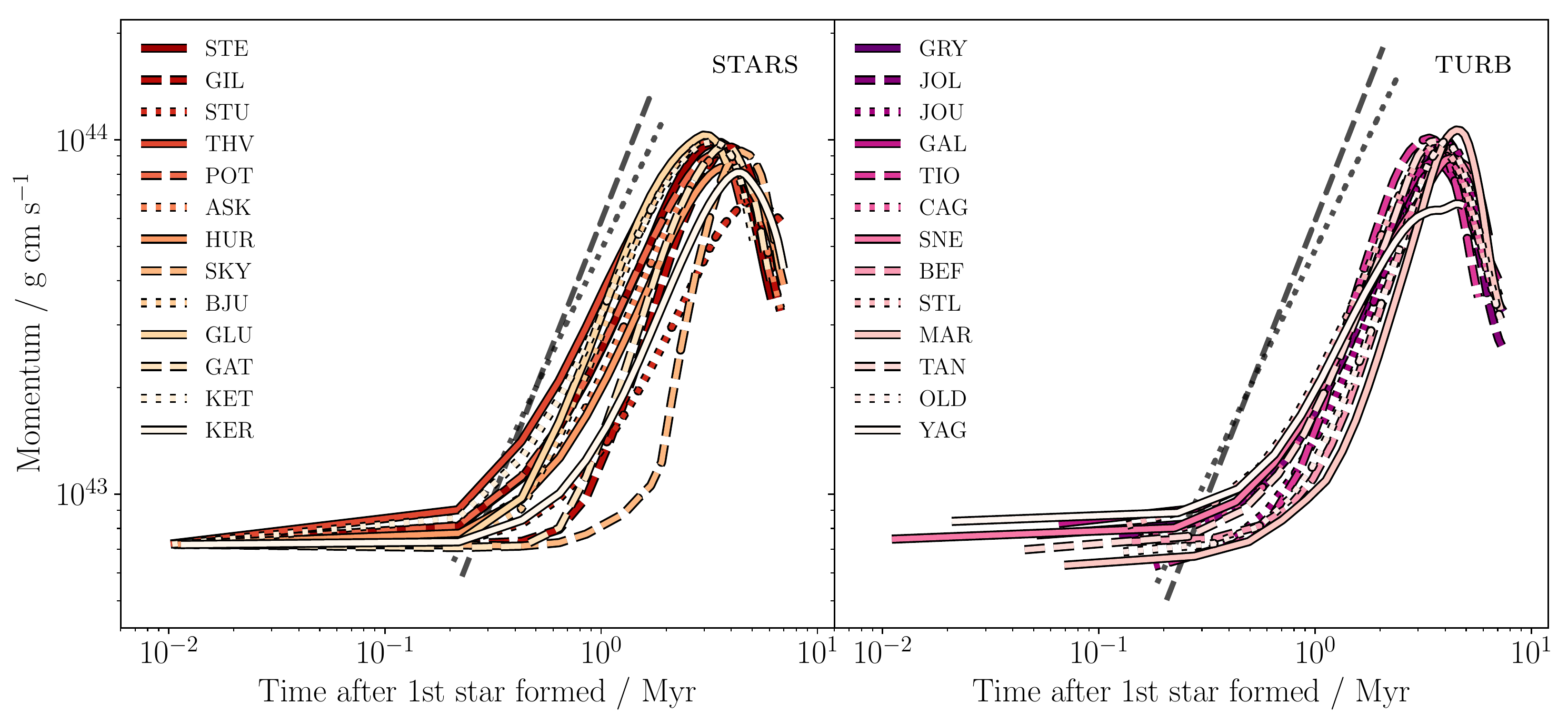}}
	\caption{Total gas momentum in the simulation volume over time for each of the simulations for flows in all directions. As in Figure \protect\ref{nphotons}, on the left are simulations in which we vary the mass of massive stars formed. On the right are simulations where the initial random seed of the turbulent velocity field is varied. Note that all simulations have some momentum from turbulence before the first HII regions drive radial outflows. The drop in momentum around $10^{44}$ g cm/s is due to flows leaving the simulation volume. We overplot the gradient of the momentum solution derived from Equation \protect\ref{powerlawradius} for a flat density profile (dotted black line) and for a profile of $w=1.71$, the power law fit to Figure \protect\ref{rayprofs} (dashed black line), in order to illustrate the time dependence of the solution.}
	\label{momentuminsnap}
\end{figure*}

All of the clusters generate considerable amounts of momentum in the cloud (up to $10^{44}$ g cm/s, see Figure \ref{momentuminsnap}) before flows leave the box and their momentum is no longer tracked. Most simulations exhibit a similar rate of momentum increase, although clusters with less efficient feedback (i.e. higher \SFE) also generate less momentum. 

Unlike the radius of the HII region discussed in Section \ref{globalprops:radius}, the analytically-derived momentum injection rate shows less dependence on the density profile, with the fits for a uniform and power law profile giving similar gradients. Looking at the solutions for a uniform density field in \cite{Matzner2002}, the radial expansion of a cloud has a weaker dependence on time and photon emission rate than momentum, but a stronger density dependence. Based on this, we suggest that momentum of ionisation fronts is more strongly influenced by the structure of the gas around it than the photon emission rate of the source, although this is important to obtain an accurate solution.

Much of this momentum is transferred to the ambient medium in the cloud. The whole box including the external medium has an initial mass of approximately $10^5$ \Msolar. $10^{44}$ g cm/s spread over $10^5$ \Msolar gives an average speed of 5 km/s. However, initially the HII region travels supersonically, since Equation \ref{powerlawradius} can give $\dot{r_i}>c_i$ if $w$ is large enough.

In this paper we do not include momentum from direct radiation pressure. The largest photon emission rate in our study is approximately $3\times10^{49}$ s$^{-1}$ (see Figure \ref{nphotons}). Assuming direct momentum deposition from photons over 4 Myr, we obtain $3\times10^{42}$ g cm/s if the average photon energy is 13.6 eV (the ionisation energy of hydrogen) or $10^{43}$ g cm/s if the average photon energy is 55 eV (the average energy in our HeII-ionising photon bin). The maximum possible momentum deposition is thus non-negligible compared to the momentum in flows in the cloud prior to the first star being formed, but is an order of magnitude smaller than the momentum of the D-type photoionisation front.

The picture so far is of HII regions that overtake their host clouds on the order of 0.3 to 2 Myr and escape into the external medium. At some point, these expanding HII regions stop the accretion onto their host clouds and freeze out the star formation to give a final \SFE. In the next Section we discuss whether or not we can identify trends in how this final \SFE is obtained.

\section{Predicting the SFE of a Cloud}
\label{correlations}

Thus far, we have established that it is difficult to accurately predict the \SFE of a cloud knowing only its initial global properties such as mass, radius or virial parameter. However, there may be some causal link between emergent properties of the cloud and cluster, and the resulting \SFE. In other words, we wish to know whether a predictive, quantitative model for the feedback loop of star formation in clouds can be formulated, or whether star formation relations are the statistical combination of events that are mostly random on cloud scales. This process is complicated by the interaction between randomised model choices (i.e. \IMF sampling and initial turbulent velocity field) and nonlinear effects that will amplify small perturbations to the system.

\begin{figure*}
	\centerline{\includegraphics[width=0.95\hsize]{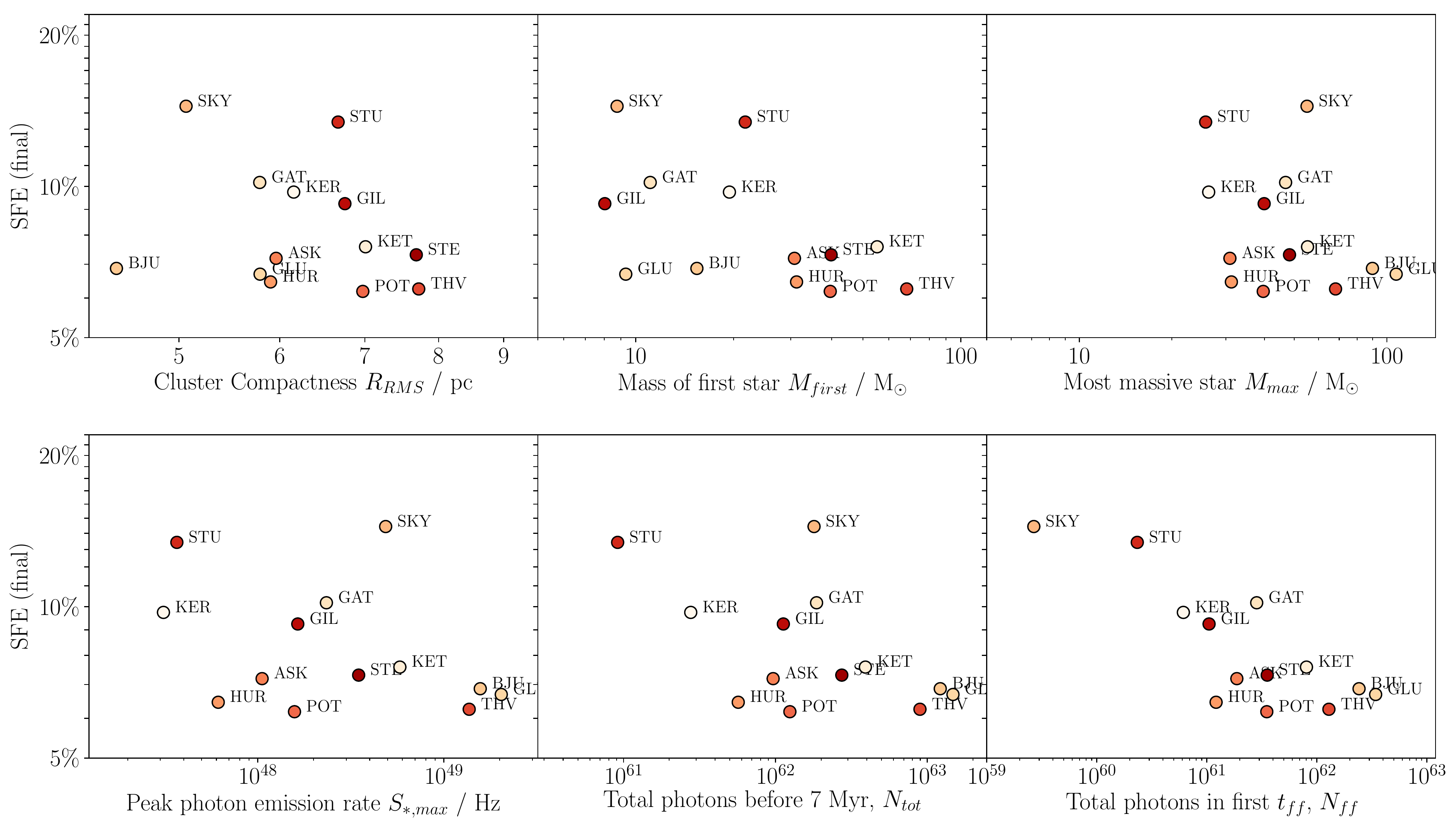}}
	\caption{Star formation efficiency (SFE) given as the final fraction of the cloud's initial mass converted to stars (i.e. sink particles), plotted against various cloud properties for each of the \protect\stars simulations listed in Section \ref{correlations:parameters}. Values are plotted on a log scale to capture the dynamic range of the various properties. Simulation code names are labelled on each of the points, while point colour corresponds to the same encoding used in previous figures. Cluster compactness is measured at $t=1.5 t_{ff}$, i.e. $t_{ff}$ after gravity is turned on in the simulation.}
	\label{starrelations}
\end{figure*}

\begin{figure}
	\centerline{\includegraphics[width=0.8\hsize]{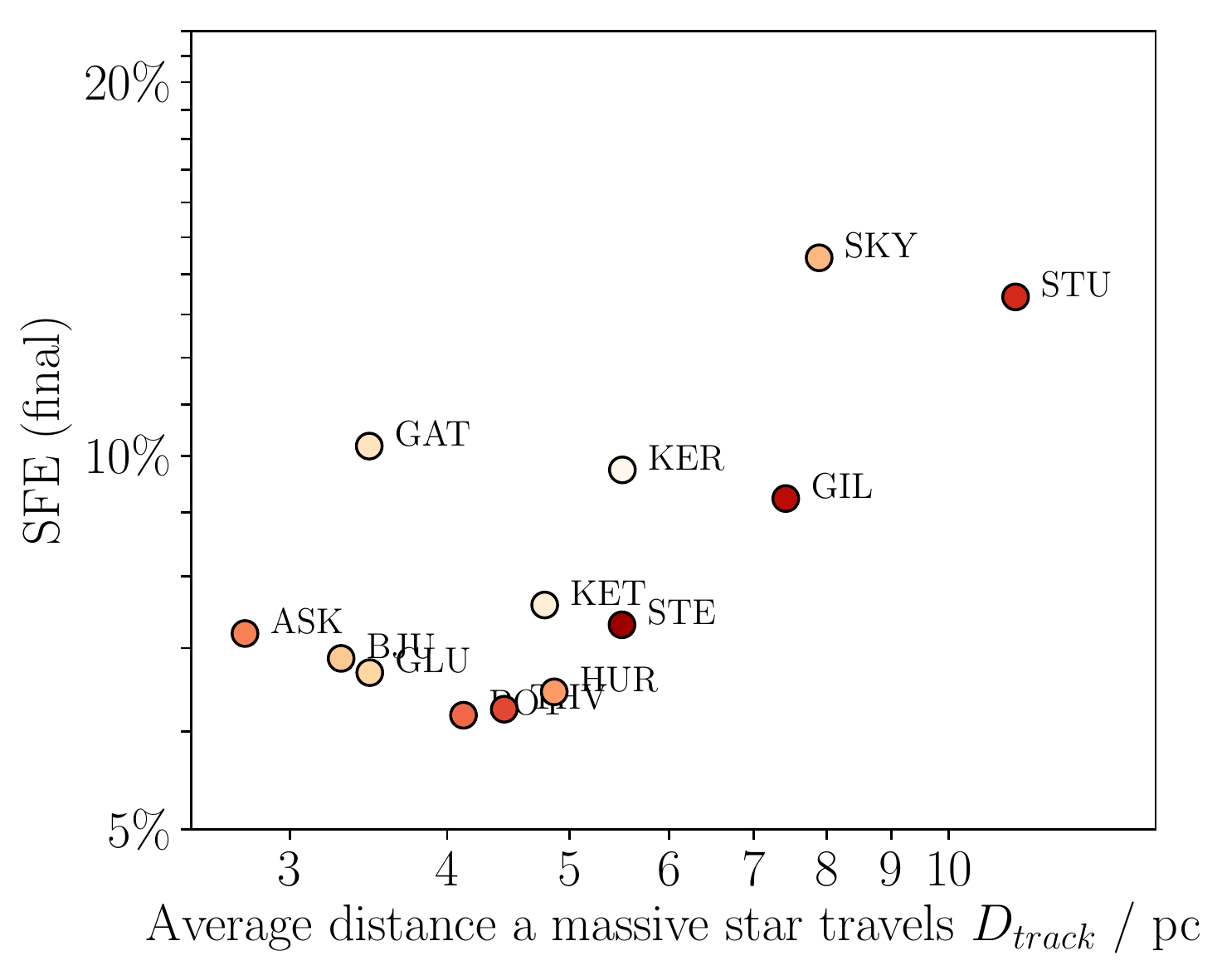}}
	\caption{Addition to Figure \ref{starrelations}, showing the average distance travelled by a massive star over its lifetime against the final SFE.}
	\label{tracklength}
\end{figure}

\begin{figure*}
	\centerline{\includegraphics[width=0.95\hsize]{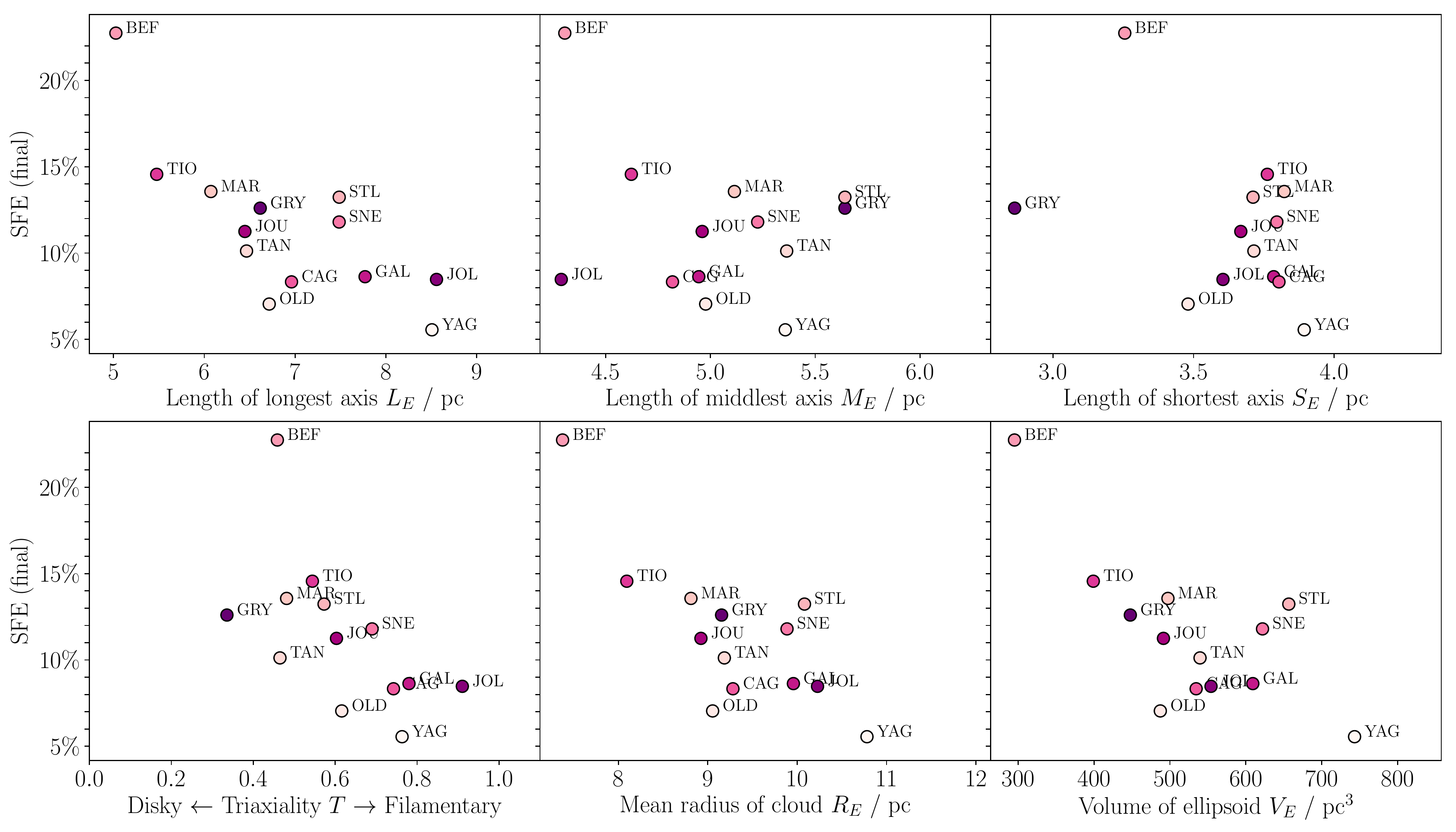}}
	\caption{Star formation efficiency (SFE) given as the final fraction of the cloud's initial mass converted to stars (i.e. sink particles), plotted against various cloud properties for each of the \protect\turb simulations listed in Section \ref{correlations}, similar to Figures \ref{starrelations} and \ref{tracklength} for the \protect\stars set. Values here are plotted on a linear scale, since the dynamic range of values is lower, and triaxiality $T$ has valid values from 0 to 1. The lower panels are quantities derived from $L_E$, $M_E$ and $S_E$. Simulation code names are labelled on each of the points. All of the values are measured with a cut-off density of 10 cm$^{-3}$ at $t = 1.5 t_{ff}$ - see Section \ref{correlations:parameters} for a discussion of these choices. BEF\protect\turbsub is more compact than the other clouds, though it lies on the same relationship between SFE and the other variables measured for all \turb simulations.}
	\label{correlatestructure}
\end{figure*}

\subsection{Parameter Selection}
\label{correlations:parameters}

To understand which parts of the system are important for predicting the \SFE, we must first identify which parameters are well correlated with the final \SFE of each cloud, and which show no correlation. For some parameters we must select a time at which the parameter is measured. We discuss the consequences of this choice of time in Section \ref{correlations:physical}. We attempt to capture a large range of cloud properties that are relevant to the global cloud structure and cluster as a whole. In future work we will revisit the dataset to quantify more detailed propreties of indivdual system components that influence the star formation feedback loop.

In the \stars set of simulations, we identify the following parameters: the most massive star ($M_{max}$), cluster compactness ($R_{RMS}$, the RMS distance of each sink particle from the centre of mass of the cluster after 1.5 $t_{ff}$, i.e. $t_{ff}$ after gravity is turned on), the mass of the first star formed ($M_{first}$), the peak photon emission rate ($S_{*,max}$), the total number of photons emitted by the cluster ($N_{tot}$), the total number of photons emitted by the cluster in the first freefall time, i.e. 0.5 $t_{ff}$ after gravity is turned on ($N_{ff}$), and the mean distance travelled by a massive star during its lifetime ($D_{track}$).

In the \turb set of simulations, we focus on the global structure of the cloud. We make an ellipsoid fit to the gas density field using the algorithm described in \textsc{Halomaker} \citep{Tweed2009a}. To do this, we include every cell above a density threshold $n_{thresh}$ and weight by the mass of the cell. We perform this fit in each simulation at 0.5, 1.0, 1.5 and 2.0 $t_{ff}$. We identify the axes of the ellipsoid $L_E$, $M_E$ and $S_E$, from longest to shortest. We then calculate the following derived structural parameters: triaxiality $T_E \equiv (1 - \beta^2) / (1 - \gamma^2)$, where $\beta \equiv M_E/L_E$ and $\gamma \equiv S_E/L_E$ \citep[see][]{Kimm2007}, mean ellipsoid radius $R_E\equiv\sqrt{L_E^2+M_E^2+S_E^2}$ and ellipsoid volume $V_E\equiv(4/3) \pi L_E M_E S_E$. 

We chose a value of $n_{thresh} = 10~$cm$^{-3}$, encompassing most of the cloud material above the background density of 1 cm$^{-3}$. Larger values led to worse correlations. Observational work \citep[e.g.][]{Lada2010} and theoretical work \citep{Geen2017} suggests that dense gas alone is a better predictor of recent star formation rather than the total star formation history of the cloud. We discuss this difference further in Section \ref{discussion:observational-significance}. We limit ourselves to the list of parameters mentioned above in the first instance, with the intention of expanding the scope of future parameter studies once we have understood the response of the system to this parameter set.

In Figures \ref{starrelations}, \ref{tracklength} and \ref{correlatestructure} we plot these parameters for each of the clouds against \SFE. It is possible to visually identify various correlations in certain parameters, and poor correlations in others, although in all cases there is scatter in the results. In the next Section we discuss a Bayesian model used to determine which parameters have strong relationships with the \SFE, and which have no clear effect on the outputs of the system.

\subsection{Setup of Model Comparison}
\label{correlations:models}

\begin{figure}
	\centerline{\includegraphics[width=0.9\hsize]{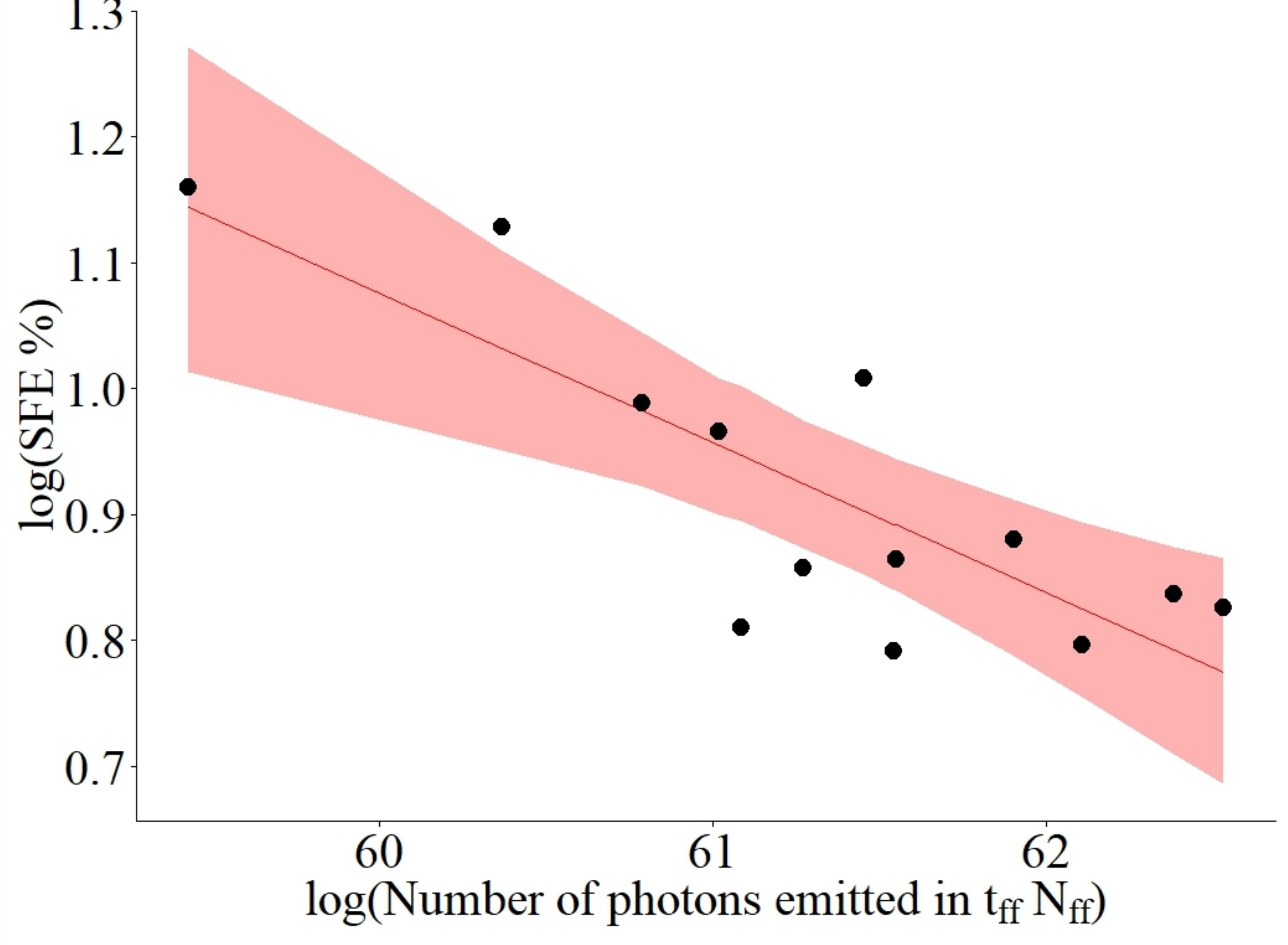}}
	\caption{Predicted values of SFE (as a percentage of the inital gas mass converted into stars) against $N_{ff}$ (see Table \protect\ref{tableX}). The red shaded area is the 95\% credible interval. Points show the sampled values in each of the \protect\stars simulations.}
	\label{figureX}
\end{figure}

\begin{figure*}
	\centerline{\includegraphics[width=0.9\hsize]{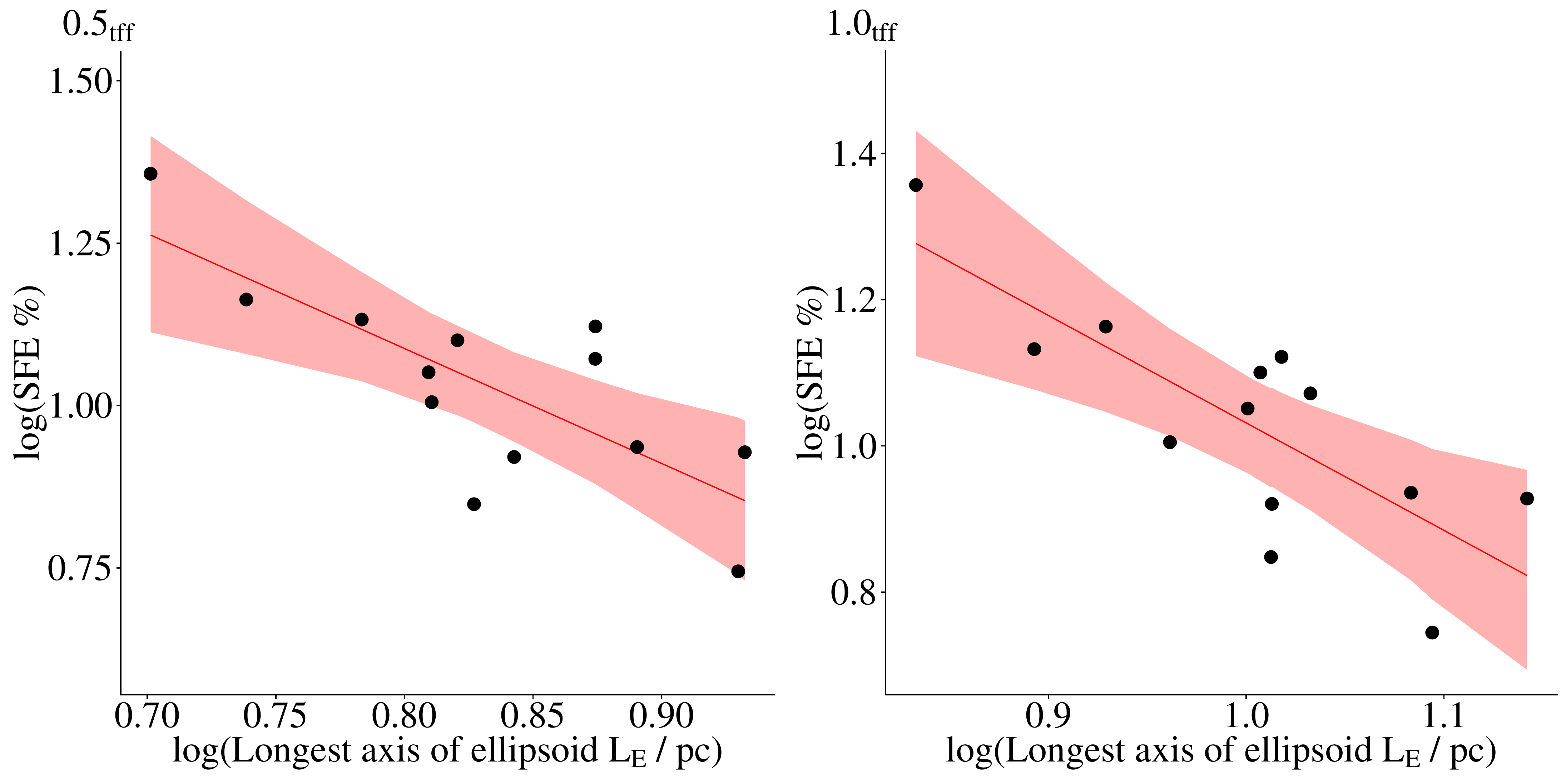}}
	\caption{Predicted values of SFE (as a percentage of the initial gas mass converted into stars) against cloud length $L_E$, measured as the longest axis of an ellipsoid fit to all gas above 10 cm$^{-3}$. Figures for $R_E$ and $V_E$ are not included since they are both dependent on $L_E$. The left panel is taken at 0.5$t_{ff}$ and the right panel at 1.0$t_{ff}$. There is a strong relationship between \protect\SFE and $L_E$ before $t_{ff}$ (and, by extension, $V_E$ and $R_E$), with little sign of time evolution in the gradient or intercept. After this time the relationship becomes weaker. The red shaded area is the 95\% credible interval. Points show the sampled values in each of the \protect\turb simulations.}
	\label{figureY}
\end{figure*}

\begin{table}
	\centering
	\begin{tabular}{llllll}
		\thead{Model \\ structure} & \thead{$\mathbf{t/t_{ff}}$} & \thead{Fixed \\ effect} & \thead{Posterior \\ mean} & \thead{Lower \\ 95\% CI} & \thead{Upper \\ 95\% CI} \\
		\hline
%		SFE $\sim 1$ &  -  & Intercept    & 6.934  & 1.786  & 11.92 \\
%		$~+ R_{RMS}$ &  1.5   & $R_{RMS}$    & 0.186  & -0.880 & 1.219  \\
%		$~+ M_{first}$ &  -   & $M_{first}$  & -0.139 & -0.372 & 0.103  \\
%		$~+ N_{tot}$ &  -   & $N_{tot}$    & 0.0226  & -0.0952 & 0.150  \\
%		$~+ N_{ff}$ &   -  & $N_{ff}$     & -0.115 & -0.208 & -0.029 \\
		SFE $\sim N_{ff}$ & - & Intercept & 8.263  & 4.567  & 12.082 \\
		                  &   & $N_{ff}$  & -0.119 & -0.171 & -0.059 \\
		SFE $\sim$ $L_E$             & 0.5          & Intercept             & 3.621                   & 2.046                  & 5.232                  \\
		&              & $L_E$                     & -2.683                  & -4.367                 & -1.056                 \\
		SFE $\sim$ $R_E$         & 0.5          & Intercept             & 3.621                   & 2.046                  & 5.232                  \\
		&              & $R_E$                 & -2.683                  & -4.367                 & -1.056                 \\
		SFE $\sim$ $V_E$         & 0.5          & Intercept             & 4.053                   & 2.060                  & 6.048                  \\
		&              & $V_E$                 & -1.114                  & -1.849                 & -0.370                 \\
		SFE $\sim$ $L_E$             & 1.0          & Intercept             & 3.247                   & 1.915                  & 4.546                  \\
		&              & $L_E$                     & -2.015                  & -3.155                 & -0.768                 \\
		SFE $\sim$ $R_E$         & 1.0          & Intercept             & 3.247                   & 1.915                  & 4.546                  \\
		&              & $R_E$                 & -2.015                  & -3.155                 & -0.768                 \\
		SFE $\sim$ $V_E$         & 1.0          & Intercept             & 3.601                   & 1.150                  & 6.048                  \\
		&              & $V_E$                 & -0.868                  & -1.683                 & -0.032                 \\
		SFE $\sim$ $T$             & 1.5          & Intercept             & 0.906                   & 0.714                  & 1.119                  \\
		&              & $T$                     & -0.704                  & -1.717                 & 0.285                  \\
		SFE $\sim$ $R_E$         & 1.5          & Intercept             & 1.582                   & 0.535                  & 2.621                  \\
		&              & $R_E$                 & -0.451                  & -1.256                 & 0.439                  \\
		SFE $\sim$ $V_E$         & 1.5          & Intercept             & 0.951                   & 0.015                  & 1.859                  \\
		&              & $V_E$                 & 0.025                   & 0.277                  & 0.314                  \\
		SFE $\sim$ $M_E$             & 2.0          & Intercept             & 1.237                   & 0.881                  & 1.556                  \\
		&              & $M_E$                     & -0.235                  & -0.608                 & 0.129                  \\
		SFE $\sim$ $R_E$         & 2.0          & Intercept             & 1.012                   & 0.271                  & 1.776                  \\
		&              & $R_E$                 & 0.011                   & -0.516                 & 0.581                  \\
		SFE $\sim$ $V_E$         & 2.0          & Intercept             & 1.252                   & 0.685                  & 1.864                  \\
		&              & $V_E$                 & 0.071                   & -0.268                 & 0.102                 
	\end{tabular}
	\caption{Summary outputs for each reported model in Section \protect\ref{correlations:models}. We determine in each case whether power laws between the input variables and SFE exist, i.e. linear relationships in log space (see Equation \protect\ref{glmmrelationship}). Each effect has a distribution of possible values, where zero indicates no relationship (i.e. the gradient between a given variable and SFE is flat). A pair of 95\% credible intervals (CIs) that are both above or below zero indicate that there is a 95\% probability that the predictor has an effect on SFE. The posterior mean is the mean value of the effect, i.e. the expected regression coefficient or intercept. We sample the cloud structure variables at various times to determine whether there is a time evolution in the fit, i.e. whether the SFE can be predicted better at earlier or later times.}
	\label{tableX}
\end{table}

To identify relationships between each of these parameters and the \SFE, we fit a series of generalised linear mixed-models (GLMMs) using a Bayesian framework and Markov Chain Monte Carlo (MCMC) methods \citep{Watson2018}. A GLMM is a form of linear regression that allows for random effects in the linear predictors that do not have to follow a normal distribution. In this analysis we assume a power law relationship between the input variables and SFE, i.e. linear relationships in log space, 
\begin{equation}
y = I + \vec{\beta}\bigcdot\vec{v},
\label{glmmrelationship}
\end{equation}
where $y\equiv\mathrm{log(SFE)}$, $I$ is the intercept, $\vec{\beta}$ is a list of fixed effects (i.e. regression coefficients corresponding to gradients in log space between SFE and each input variable, or power law indexes in linear space for each model) and $\vec{v}$ is a list of predictors or input variables in log space, such as the number of photons emitted in $t_{ff}$ ($N_{ff}$) or the length of the ellipsoid fit to the cloud ($L_E$). Each effect has a distribution of possible values, where zero indicates no relationship (i.e. the gradient between a given variable and SFE is flat). The posterior mean is the mean value of the effect, i.e. the expected regression coefficient or intercept. The credible intervals (CIs) are here equivalent to the confidence interval, and are discussed below. See \cite{VandeSchoot2014} and \cite{Harrison2018} for an introduction.

The GLMMs are computed using `RStudio' \footnote{https://www.rstudio.com/}, a development environment for the statistical computing language `R'\footnote{https://www.r-project.org/}, with the package `MCMCglmm' \citep{Hadfield2010}. MCMC chains were run for $10^5$ iterations, with a burn-in period of 1000 iterations and a thinning interval of 10 iterations. The burn-in period is the number of iterations before any results can be sampled, to reduce the influence of the random starting point of the sampling process. The thinning interval is the interval between which samples are rejected to reduce autocorrelation between samples. All models are fit with uninformative priors, i.e. making no prior assumptions about the response of SFE to the predictor variables. To this end, we use a univariate inverse Wishart distribution with $V = 1$, $\nu = 0.002$, which approximates to a delta function at 0. All models use a Gaussian distribution and identity link function, i.e. using a normal distribution to link the predictor variables to the model output. Model convergence is assessed visually using trace plots of posterior distributions and acceptably low levels of autocorrelation are achieved by ensuring that all estimated parameters have an effective sample size of over 5000.

We examine Variance Inflation Factors (VIF) to determine whether predictors have high collinearity and remove them accordingly. The VIF is the ratio of variance in a model with multiple terms and the variance of a model with one term only. A high VIF means that a parameter can be linearly predicted from another parameter with high accuracy, and is thus unlikely to be an independent variable. In the \stars analysis, three variables ($M_{max}$, $N_{TOT}$ and $S_{*,max}$) have a VIF of greater than 10 (since they can be calculated directly from the stellar evolution model for a fixed set of stellar masses) and are therefore removed from analysis. VIF for all variables in the \turb analysis are below 10. 

For each analysis we first run a `Full' model containing all possible fixed effects (variables that \SFE is thought to be dependent upon), a `Null' model containing no fixed effects and a suite of models containing each possible combination of fixed effects to determine which configuration best predicts the data. To select the best fitting model, we take an information-theoretic approach \citep[see][]{Burnham2003}, meaning that a deviance information criterion (DIC) is computed from each model and compared. The DIC is a measure of quality for a given model, which maximises the probability that a given model is correct using the maximum likelihood estimate for a set of parameters, while minimising the number of parameters included in the model. A lower DIC indicates a better fit for the data. Where there is no clear best fitting model, the most parsimonious (fewest predictors) model with the lowest DIC is chosen. In this paper we use the \cite{Spiegelhalter2002} definition for DIC. Whether a given model fits the data better than another is determined by whether there is a substantial difference ($>$2) in DIC between competing models. See Table \ref{tableX} for a summary, and Appendix \ref{appendix:morestats} for model outputs and DIC values from the Full, Null and Best Fitting model for each analysis.

In the \turb case, we carry out two further analyses. In the first, we fit a model using only $R_E$ as a predictor. In the second, we fit a model using only $V_E$ as a predictor. These factors could not be included in the previous analyses because they are derived from $L_E$, $M_E$ and $S_E$, which are already included in those models. Fixed effects are determined as having an important influence according to whether the 95\% credible intervals (`95\% CI') of their posterior distribution crossed zero. If a variable has a negligible effect, we expect its posterior distribution to be centred close to zero. An influential variable is expected to be shifted away from and not substantially overlapping zero (where zero indicates a flat gradient, i.e. the variable has negligible influence on the SFE).

\subsection{Results of Model Comparison}
\label{correlations:modelresults}

In the \stars analysis, the best fitting model contains only $N_{ff}$ (the number of photons emitted in the first $t_{ff}$). There was good evidence that $N_{ff}$ has a negative relationship with SFE (see Table \ref{tableX}, Figure \ref{figureX}). %(posterior mean: -0.119, 95\% CI [-0.171, -0.059], Figure X).
We do not include the mean distance travelled $D_{track}$ in the model. This is because it is an output parameter of the simulation rather than an initial condition such as the cluster properties or gas distribution. However, $D_{track}$ and \SFE do show reasonable correlation, which we discuss in Section \ref{correlations:tracklength}.

We perform the \turb analysis at a number of times to determine whether the evolution of the cloud affects the ability for the model to recover an SFE. At 0.5 and 1.0 $t_{ff}$, the best fitting models contain only $L_E$ as a predictor and there is good evidence that this variable has an important effect on \SFE (see Table \ref{tableX}, Figure \ref{figureY}). At 0.5 and 1.0 $t_{ff}$, in their respective models $R_E$ and $V_E$ are also found to have robust effects on \SFE, since they are dependent on $L_E$. At 1.5 and 2.0 $t_{ff}$, there is no strong evidence of an effect of any parameter in the \turb set on the \SFE.

We cross-check these results with a frequentist approach using the Pearson correlation coefficient and found reasonable agreement. Variables identified as having an important effect on the \SFE in the Bayesian analysis have a Pearson correlation coefficient with a magnitude of at least 0.7 .

\begin{figure}
	\centerline{\includegraphics[width=0.98\hsize]{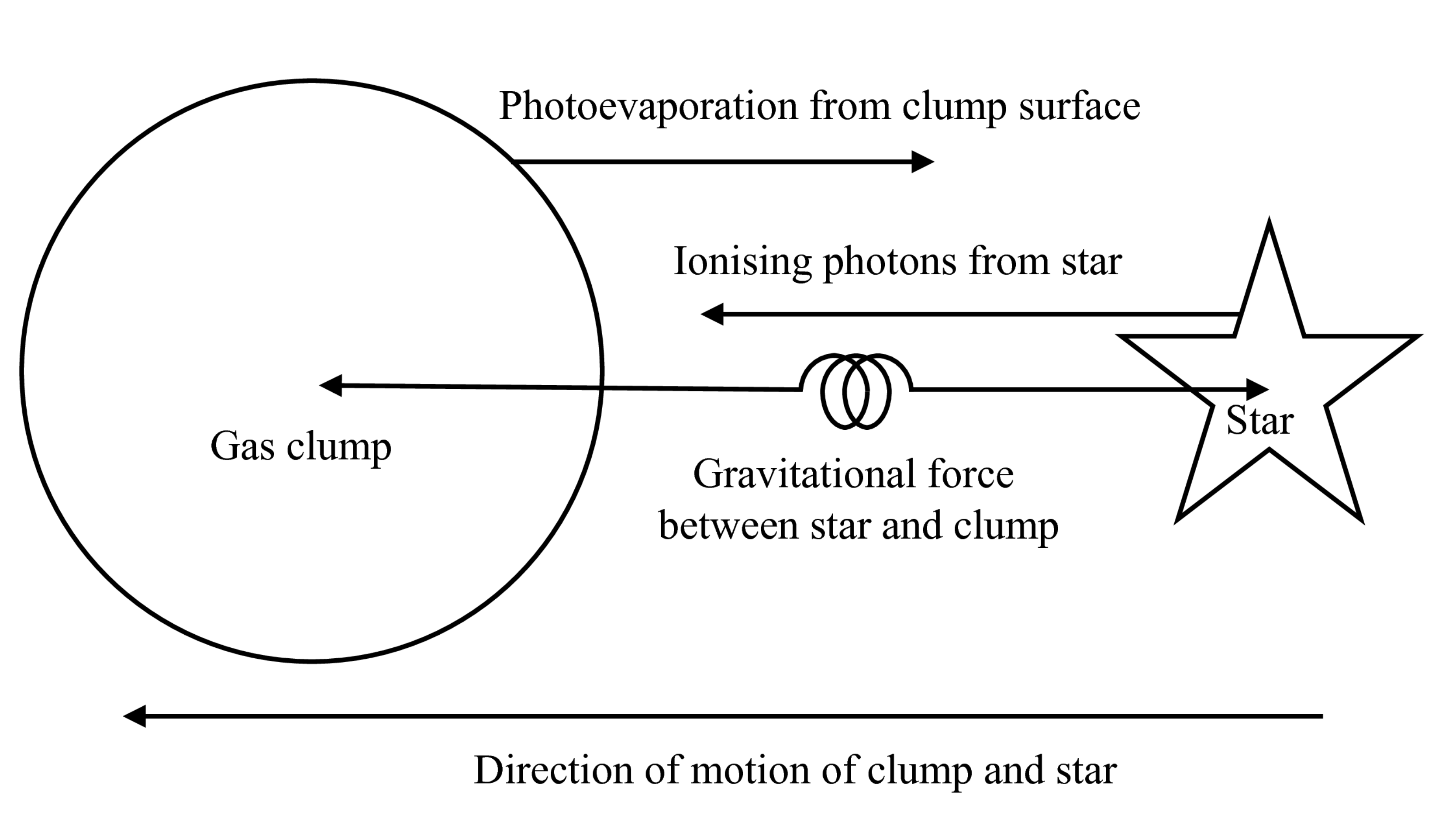}}
	\caption{Cartoon of clump-star acceleration process described in Section \protect\ref{correlations:physical}. Ionising photons from the star evaporate gas from the surface of the nearby clump facing the star (rightwards in the cartoon). This evaporation pushes the clump left in the cartoon via the rocket effect \protect\citep{Oort1955}, with the system as whole conserving momentum. If the gravitational force between the star and the clump is strong enough, the system remains bound, and the star-clump system is accelerated towards the left of the image until the clump is completely evaporated or the rocket acceleration exceeds the gravitational acceleration between the star and clump.}
	\label{clump_cartoon}
\end{figure}

\begin{figure*}
	\centerline{\includegraphics[width=0.88\hsize]{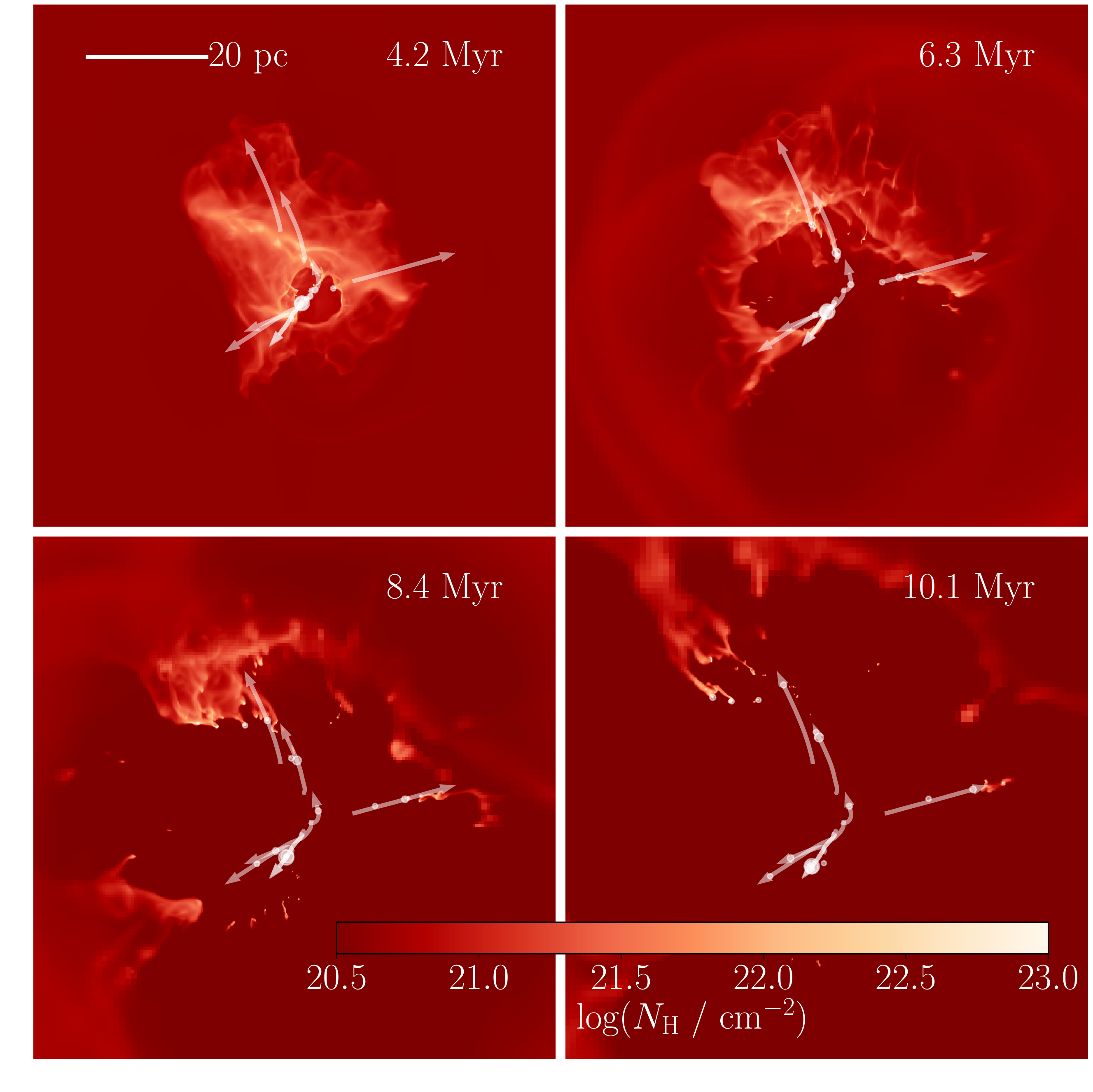}}
	\caption{Sequence of hydrogen column number density $N_H$ maps STU\protect\starsub simulation, demonstrating the motion of massive stars and nearby massive gas clumps moving outwards from the site of star formation. Each image is 78 pc across, zoomed into the central 70\% of the full box size (112 pc). White dots show the location of sink particles, with dot size proportional to sink mass. White lines show the track followed by each sink particle containing a massive star during the star's lifetime, with arrows at the end of each track. This is a highly simplified picture, which also contains effects from other stars and the shell around the HII region.}
	\label{driftvis}
\end{figure*}

\subsection{Physical Significance}
\label{correlations:physical}

Of all the parameters derived from the \IMF sampling (the \stars analysis), only the number of photons emitted in the first $t_{ff}=4.2$ Myr (i.e. 0.5 $t_{ff}$ after the gravity is first turned on) is a good predictor of the final \SFE of the cloud. In Figure \ref{tsfe}, most of the star formation in the \stars set (left panel) has happened before this time. Any further photon emission affects only residual star formation, or the ability of the cloud to recollapse and form more stars \citep{Rahner2017a}.

In the \turb analysis, clouds that are more compact have a higher \SFE (see Figure \ref{figureY}). It is curious that the longest axis $L_E$ is a better predictor than the other axis lengths $M_E$ or $S_E$. One explanation is that our clouds are, by visual inspection and by looking at the distribution of triaxiality $T$, often filamentary (or close to spherical) rather than disky. This is supported by observations, where \cite{Konyves2015a} find that 75\% of pre-stellar cores are in filamentary structures. For a fixed mass, longer filaments will have a lower concentration assuming that the width does not also strongly vary.

The fact that there is no clear relationship between the \SFE and global cloud properties at later times (1.5 $t_{ff}$ and later) adds weight to the argument that the early evolution of clouds is most important in setting the star formation properties of the system. We discuss the significance of our results to observed measures of \SFE in Section \ref{discussion:observational-significance}.

\subsection{Distance Travelled by the Stars}
\label{correlations:tracklength}

In Figure \ref{tracklength}, there is a link between the \SFE and the distance travelled by the stars $D_{track}$. This suggests that when feedback is inefficient (i.e. \SFE is high), stars travel further from the point at which they are formed. Figure \ref{imfvis} supports this argument, since simulations that at $1.5 t_{ff}$ retain some amount of dense gas close to the cluster also show the longest distances over which the stars travel, shown as white lines.

\cite{Gavagnin2017} find this effect and argue that it is due to N-body interactions between stars. Our mass resolution in this study is not sufficient to resolve the full stellar cluster N-body dynamics. Instead, as in \cite{Geen2017}, we offer the following explanation. The star produces ionising radiation that evaporates material from the gas clump. This causes the gas clump to accelerate away from the star via the rocket effect \citep{Oort1955}. However, since the star and gas clump are gravitationally bound, the star also moves in the same direction. We illustrate this process with a cartoon in Figure \ref{clump_cartoon}. This continues until the gas clump is completely evaporated or the rocket acceleration exceeds the gravity between the star and gas clump.

We visualise this process in the STU\starsub simulations in Figure \ref{driftvis}, which has one of the longer track lengths in Figure \ref{tracklength}. Massive stars accelerate over their lifetimes in the direction of the nearest gas clumps. The second generation of stars form in clumps already moving away from the centre of the cloud due to HII regions from the first generation of stars, but this initial velocity accounts for only $\sim$10\% of the distance travelled by the stars. This picture is greatly simplified, since there is a complex interaction between a given clump-star pair and other stars and gas structures in the cloud.

This gives the counterintuitive result that, in the low \SFE regime studied in this paper, inefficient feedback causes the stellar cluster to disperse, while efficient feedback that quickly removes most of the cloud material causes the cluster to remain closer to where it was formed. We do not resolve the full N-body dynamics of the cluster, since we do not form individual stars with masses below 1 \Msolar, so we cannot comment on whether this would alter our results.

\section{Discussion}
\label{discussion}

In this Section we discuss the significance of our results, and how they should be interpreted in a wider context.

\subsection{How Predictable is the SFE?}
\label{discussion:howpredictable}

This paper suggests that the \SFE of clouds similar to those in the solar neighbourhood \citep[by comparison of column density distributions in][]{Geen2017} is difficult to predict. This makes a pure simulation approach to determining the final \SFE of clouds across a wider parameter space of \ISM properties prohibitively expensive, since the full parameter space of the problem requires multiple realisations to capture the range of values relevant to each stochastic process.

To first order, the \SFE is closely linked to how many ionising photons are emitted during the peak of star formation and how compact the cloud structure is. This is, at first glance, in agreement with existing models that include feedback processes that end star formation in a cloud \citep[e.g.][]{Matzner2002,Krumholz2006,Goldbaum2011}. However, it is clear from this work and others that the clouds are more complex than a spherical system with a central point source representing an accreting cluster, as many analytic models assume. While we recover some trends in star formation efficiency, scatter remains in our results, showing that the systems are not completely predictable, even when randomised model inputs are well constrained.

We have established that the ionising photon emission rate at early times is linked to the \SFE. In turn, the \SFE sets the number of massive stars produced. Since our cloud mass is fixed at $10^4$ \Msolar, we form a new massive star every time the SFE increased by 1.2\%. Our median \SFE in the \stars runs was 7.3\%, which equates to 6 massive stars. This means that there is a large Poisson sampling error on the results (the Poisson error for 6 samples is $\sim40$ \%). As shown in Figure \ref{qdispersion}, the stochastic effect of \IMF sampling will become smaller in more massive clusters where the sampling of the high end of the \IMF is more complete. For more massive clusters, \cite{Grudic2018} find a relatively clean correlation between initial surface density and \SFE.

In this paper we use a statistical model to understand to first order which properties of a cloud affect its evolution. A further stage will be to understand in more detail how these parameters feed into more predictive models to explain at what point star formation is terminated by feedback in molecular clouds. We have shown that existing theory for the expansion of HII regions in idealised conditions can be applied to explain more complex cloud systems, at least for simple aspects such as the expansion rate of and momentum injection from HII regions in non-uniform clouds.

\subsection{Dispersion of Ionising Photon Emission Rates}
\label{discussion:emissionrates}

\begin{figure}
	\centerline{\includegraphics[width=0.90\hsize]{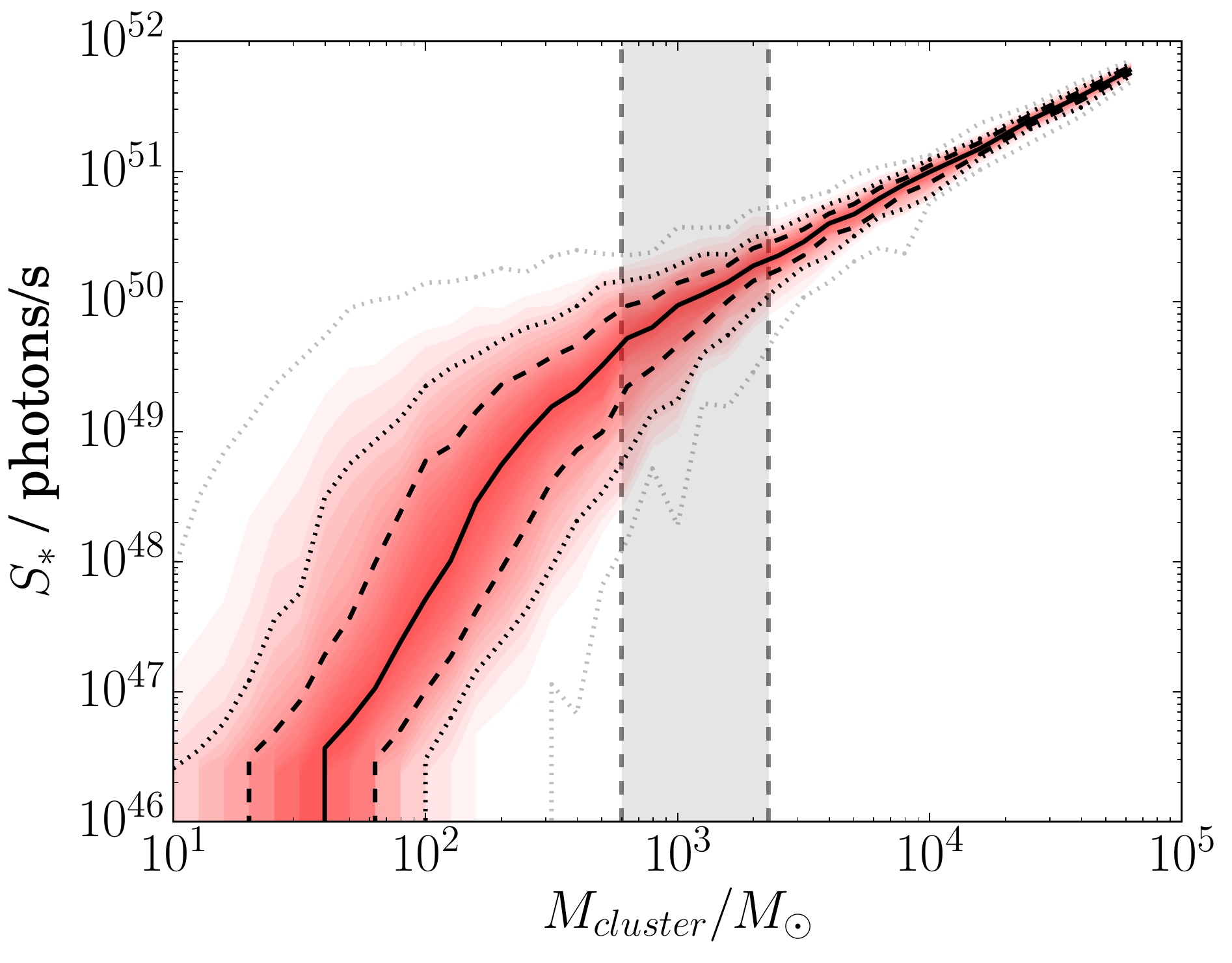}}
	\centerline{\includegraphics[width=0.90\hsize]{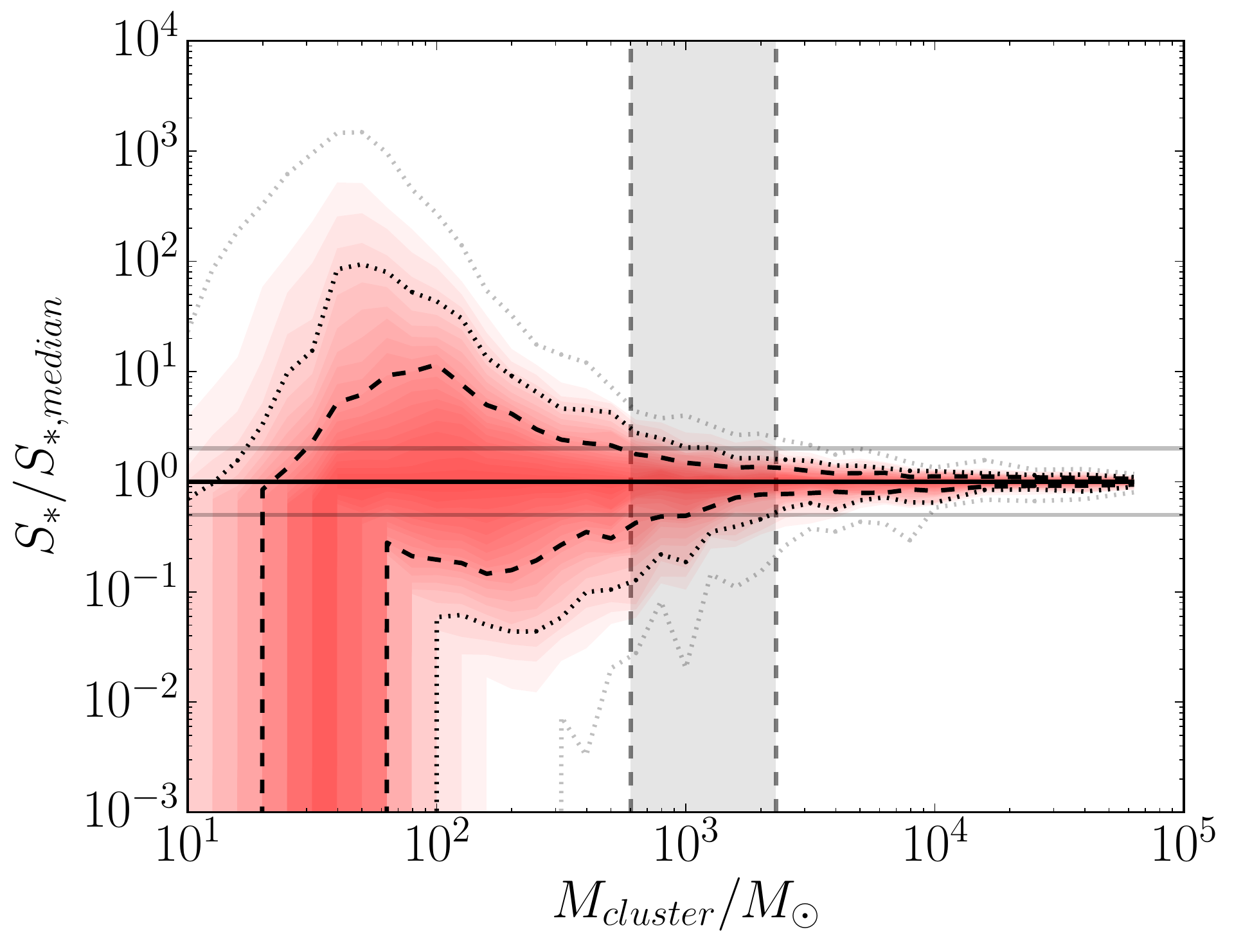}}
	\caption{The distribution of hydrogen-ionising photon emission rates $S_*$ from stochastically-sampled clusters of different masses. In each cluster mass bin we generate $N_{cluster}$ clusters by stochastically sampling a Chabrier \protect\IMF until its mass reaches the desired cluster mass. Since there is more scatter at lower cluster masses, we generate more clusters for lower masses. To this end, $N_{cluster}$ is chosen heuristically using a step function where $N_{cluster}=100$ if $M_{cluster} > 500$ \protect\Msolar, $N_{cluster}=500$ if 200 \protect\Msolar $< M_{cluster} < 500$ \protect\Msolar, and $N_{cluster}=1000$ otherwise. We calculate $S_*$ in a cluster by summing over all stars using a fit to \protect\cite{Sternberg2003} for Class V / main sequence stars. The top panel shows the total number of photons emitted by the cluster, and the bottom panel the number of photons divided by the median number of photons for the mass bin, $S_{*,median}$ (for values of $M_{cluster}$ below 50 \protect\Msolar, we use $S_{*,median}$ at 50 \protect\Msolar). The grey shaded area between the vertical dashed lines shows the range of final cluster masses sampled by this work. The grey dotted lines show the maximum and minimum $S_*$ in each mass bin. The black dotted lines show the 10\% and 90\% probability thresholds. The black dashed lines show the 25\% and 75\% probability thresholds. The median is plotted as a solid black line. Forty intermediate probability bins are plotted using red shading, with deeper red being closer to the median. In the bottom panel, the grey solid lines show 50\% and 200\% of $S_{*,median}$.}
	\label{qdispersion}
\end{figure}

Stars form with masses distributed according to an \IMF. In this paper we randomly sample a Chabrier \IMF 13 times to generate our \stars set of simulations. In Figure \ref{qdispersion} we provide some context for how this random sampling affects the ionising photon emission rate $S_*$. Here we generate 100 to 1000 clusters for each cluster mass bin $M_{cluster}$, and calculate $S_*$ for the cluster as a whole. Here, $M_{cluster}$ is the total mass of a cluster of stars. We do this by calculating $S_*$ for each star using the values given in \cite{Sternberg2003}, assuming that all stars are Class V (main sequence), and summing over the whole cluster.

We find a double power law relationship in the median $S_*$ against $M_{cluster}$. This turnover occurs around 1000 \Msolar, and happens when the probability for the maximum stellar mass $m_{max}$ ($=120$ \Msolar) exceeds 0.5. The slope below 1000 \Msolar thus depends heavily on the maximum stellar mass sampled, while the slope above 1000 \Msolar is a linear relationship following the emission rate for a well-sampled population of stars. This turnover in $m_{max}$ is also found in \cite{Weidner2009} and \cite{DaSilva2012}, who calculate the maximum mass of stars in a cluster of a given mass, although there is some dependence on the \IMF used. 

A key difference between our model assumptions and \cite{Weidner2009} is that for clusters between 1000 and 4000 \Msolar the latter paper does not find many stars above 25 \Msolar, suggesting a physical link between the distribution of the \IMF and the host gas reservoir. Since we cannot resolve the full \IMF ab initio in our simulations while producing a large sample of clouds, we cannot comment directly on whether this has a statistical or physical origin. In addition, we cannot comment on whether there is indeed a maximum stellar mass (for example, star R136a1 has a mass over 300 \Msolar \citep{Crowther2016}, although such stars are rare). A closer examination of the link between the gas reservoir, the \IMF and ionising radiation from massive stars must be left to future work.

In this study we find values for the final \SFE ranging from 6 to 23\%, or $M_{cluster}=600$ to 2300 \Msolar, which we overplot on Figure \ref{qdispersion}. On the bottom panel of this Figure we overplot 50\% and 200\% of the median $S_*$ for each mass bin. At the lower mass end, there is a factor of 4 range of values in the 25\% to 75\% limit, with more convergence at higher masses. Precisely which cluster mass the simulation ends at depends on a complex interaction between the processes of gas accretion onto sinks and the dispersal of dense gas by HII regions driven by ionising photons, and the steep profile of $S_*$ in this mass regime makes accurate \SFE predictions difficult.

% from Programming/Astro/SFE

\begin{figure*}
	\centerline{\includegraphics[width=0.48\hsize]{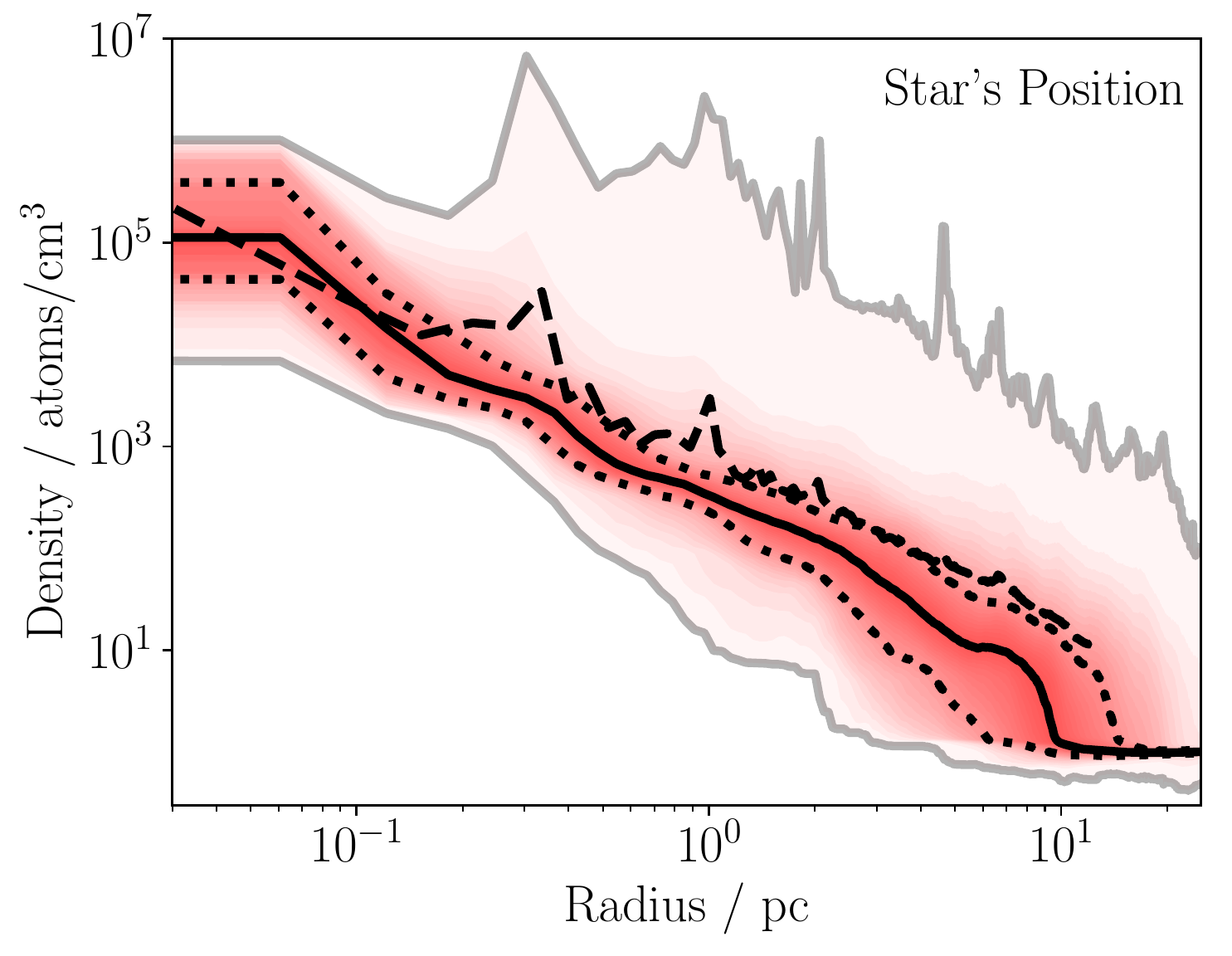} \includegraphics[width=0.48\hsize]{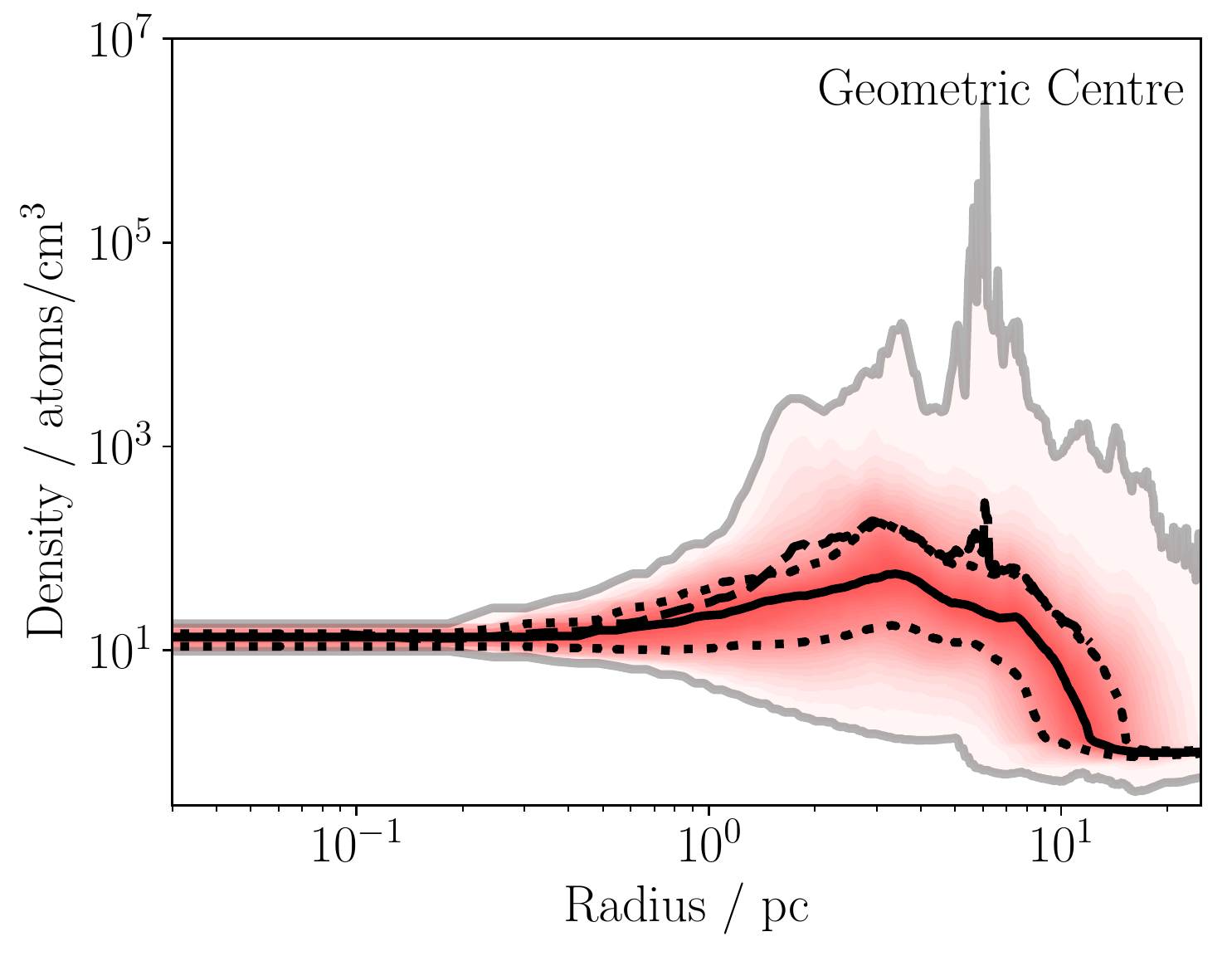}}
	\caption{The radial density distribution in the cloud at the time at which the first star forms in the \protect\stars simulations, 3.38 Myr. On the left is the profile centred on the sink particle hosting the first star to form. On the right is the profile centred on the geometric centre of the simulation volume. To generate the profiles, we cast $10^4$ rays through the volume with 1000 radial bins. The solid black line shows the median density in each radial bin. The dotted black lines show the 25\% and 75\% percentiles. The grey lines show the maximum and minimum densities at each radius. The red shading is applied in bands of 2\%, with redder values being closer to the median. The dashed black line is the mean density profile, measured by binning $10^7$ points uniformly sampled inside 12 pc around the origin. Note that our maximum spatial resolution is 0.03 pc, and values close to this radius suffer from low number statistics.}
	\label{rayprofs}
\end{figure*}

\subsection{Comparison to Systemic Variations in Cloud Properties}
\label{discussion:systemicproperties}

\rev{We should caution that in our analysis, we separate the IMF sampling and initial cloud structure as variables, and limit our analysis to a single cloud mass with global properties similar to clouds in the solar neighbourhood. In previous studies of photoionising feedback, cloud mass has been found to influence properties such as the \SFE, cluster unbinding \citep[e.g.][]{Dale2012,Dale2013} and UV escape fraction \citep[e.g.][]{Howard2018}. For example, in \cite{Dale2012}, a very dense, massive cloud achieves a \SFE close to 100\%, since photoionisation feedback becomes less effective. In most other cases, they find a \SFE close to 10\%, and argue for a window in cloud surface density where photoionisation feedback sets the final \SFE of the cloud.}

\rev{Protostellar jets can also reduce the short-term \SFE of a clump by returning gas back into the cloud. \cite{Federrath2014} find a 25\% reduction in the \SFE of a clump when jets are included, although this material is usually recycled back into the host cloud rather than being ejected from the system entirely.}

\rev{In clouds formed by accretion from larger volumes, \cite{Vazquez-Semadeni2010} find an \SFE from a few percent to 20\% when feedback is included, which is somewhat consistent with our result, where we introduce random variation by hand into isolated clouds.}

\rev{\cite{Walch2011} argue that low metallicity galaxies have low \SFR and also exhibit more bursty star formation, with cold clump survival influenced by chemistry.}

\rev{\cite{Federrath2013} study the \SFE of turbulent boxes by varying the form of the turbulence used. Since they do not include any processes where star formation is frozen out by feedback, they instead study the properties of clouds with a given \SFE at a certain point in time, which allows comparison with observed clouds in a single state. They explore a range of 0\% to 20\% \SFE, with noticeable differences to the density PDFs when magnetic field strength, Mach number and initial compressive vs solenoidal turbulence ratios are varied. This is in agreement with analytic models of star formation rates given in, e.g., \cite{Hennebelle2008}, \cite{Federrath2012}, \cite{Salim2015} and \cite{Burkhart2018}.}

\rev{It is difficult to compare our results directly to other authors since there are many systematic differences between the simulation setups and theoretical models used. A well-constrained comparison study would be needed to determine the role played by, e.g., differences in simulation methods (SPH, grids, moving mesh), physics included, resolution, etc, as well as systematic differences in the bulk properties of the cloud. However, it seems clear that where feedback is effective \citep[see][]{Dale2012}, the \SFE freezes out at anywhere between a few to a few tens of percent, roughly on the order of our results.}

\subsection{Internal versus External Processes}
\label{discussion:internalvsexternal}

The question of how clouds are formed and their relationship to the host galaxy as a whole is an important one to address. \cite{Rey-Raposo2015} have already raised the issue of cloud geometries, arguing that clouds extracted from a simulation of a galactic disk behave differently to isolated spheres. The importance of choice of initial turbulence has also been demonstrated by \cite{Goodwin2004a} and \cite{Girichidis2011,Girichidis2012a,Girichidis2012b}. Our results clearly show a link between \SFE and cloud geometry, with more elongated clouds producing fewer stars, although we cannot comment on how the galactic origin of the clouds affects their subsequent evolution. \cite{Ibanez-Mejia2017} and \cite{Seifried2018} recently argued using simulations of kpc-wide chunks of a galactic disk that external influences, such as supernovae and tidal fields, are less important than internal processes. \cite{Dobbs2013} found that spiral arms play an important role in cloud formation, though.

\subsection{Observational Significance}
\label{discussion:observational-significance}

We discussed in \cite{Geen2017} the relationship between the theoretical definition of \SFE (the total amount of gas in a cloud converted into stars), and the observational definition (the ratio of recent star formation versus current gas mass above some column density threshold). This is significant because there is no clear link between the two definitions. An analysis of observable quantities in synthetic observations of the \textsc{yule} simulations is left for future work. However, it will be valuable to extend our analysis to a comparison with observed clouds in order to obtain better statistics that allow us to interpret resolved star-forming regions in the Milky Way and Magellanic Clouds.

Observational studies typically measure \SFE as the recent star formation in a system rather than an idealised total. \cite{Lada2010}, \cite{Heiderman2010}, \cite{Gutermuth2011} and others establish some link between local density peaks and recent star formation, with weaker or no correlation to global cloud properties. In selecting parameters in the \turb set of simulations for the statistical model in this study, we found a better relationship with total \SFE when we sample the whole cloud using a low density cut-off than when we sample only the densest parts of the cloud using a higher density cut-off. One possible explanation is that star formation has trends on both short and long timescales, and that the total \SFE of a cloud is better correlated to large-scale cloud properties, while recent star formation is more closely linked to small-scale structures with shorter freefall times.

\rev{In \cite{Geen2017}, we found that the observationally-derived \SFE as defined by \cite{Lada2010} varies in a single cloud identical to the \stars simulations from 8 to 30\%, a little larger than the observed variations in the nearby Gould Belt. We stress, however, that this measurement has no link to the total \SFE of the cloud studied in this work and indeed varies with time, since the measurement considers only stars younger than 2 Myr and only gas above an extinction threshold of $A_k=0.8$.}

\subsection{Galactic Context}
\label{discussion:galactic-context}

Molecular clouds are embedded in the \ISM of a galaxy, and the internal processes of these clouds set both the \SFE and the amount of radiation and kinetic flows from the cloud as a function of time. The combined \SFE of clouds in turn sets the \SFE of a galaxy as a whole. The interface between molecular clouds and a galactic \ISM regulates a number of additional quantities, such as galactic wind rates and ionising photon escape fractions. Certain problems, such as the epoch of reionisation, can be very sensitive to the precise behaviour of unresolved structures in cosmological galactic simulations \citep[see, e.g.,][]{Rosdahl2018}, since both high escape fractions and high ionising photon emission rates are required simultaneously.

Feedback propagating inside molecular clouds follows an analytic solution described by a power law density field, since the star ``sees'' this distribution as it sits on a density peak. This has two consequences. Firstly, ionisation fronts expand more quickly in a power law density field than in a uniform medium, even exceeding the sound speed in the ionised gas. Secondly, however, the initial density that the front expands into is higher than the average density of the cloud. In denser clouds, this may make it harder for ionisation fronts to expand if they are strongly countered by accretion onto the protostellar cores.

The SFE of our clouds has a large scatter, making na\"ive predictions difficult. On a Galactic scale, the large number of such clouds will smooth the distribution of \SFE \citep[e.g.][]{Kruijssen2014,Leroy2016}, although an accurate prediction of this distribution is still required. With $\sim10$ simulations in each sample, we are able to predict the efficiency of star formation given a set of initial conditions, but we cannot comment on the role of various issues such as subgrid model choices. We have also shown that some correlations can be uncovered when certain emergent parameters are measured, such as cloud geometry or photon emission rates early in the cloud's lifetime. Further research into these correlations will allow us to develop a clearer picture of how the parameter space of cloud-scale physics can be used to explain galaxy-scale star formation rates and feedback efficiencies.

\subsection{Limiting Numerical Choices}
\label{discussion:limits}

As always, we must make certain limiting choices in order to fit the problem into available computational resources. In this study, we emphasise the need for repeatability to provide statistics on cloud behaviour, which is not possible if only one simulation can be completed with the available resources. Nonetheless, we can speculate how our results should change with a more complete physical model.

This paper does not attempt to reproduce the \IMF directly in the mass distribution of sink particles, as in, e.g., \cite{Bate2012}. The advantage of this is that we are able to arbitrarily change the sampled \IMF to test model predictions. However, there are limitations to this approach, particularly if we wish to explore whether clump evaporation by radiative feedback affects the \IMF in the cloud. \cite{Gavagnin2017} suggest that ionising radiation does indeed suppress the high mass end of the \IMF. One important question is when massive stars form compared to other stars. This is still an open question, although \cite{Cyganowski2017} find observational evidence of stars of all masses forming at similar times.

Another simplification is the lack of stellar winds. Winds, with a characteristic velocity of 1000 km/s or more, compared to 10 km/s for photoionised flows, significantly reduce the simulation timestep and thus increase the computational demand on the project. They can transfer significant momentum to the \ISM \citep[see, e.g.][]{Fierlinger2015a}, but the interaction between winds and radiative processes is complex \citep[e.g.][]{Capriotti2001,Rahner2017}. \cite{Dale2014} suggest that for the conditions of star formation considered here, photoionisation dominates over winds on a cloud scale. On a kpc scale, \cite{Peters2017} find that the combined effect of radiation, winds and supernovae causes a slightly lower star formation rate than when only winds and supernovae are included (in all cases including supernovae), although they do not discuss a case where only radiation is considered. \rev{In controlled studies around individual stars, \cite{Haid2018} find that stellar winds dominate the HII region only in conditions where the gas is already heated to $\sim 10^4$ K, such as in the warm interstellar medium (WIM).}

In addition, we omit radiation pressure. For more massive clusters, analytic models by, e.g., \cite{Rahner2017} argue that winds, supernovae and radiation pressure dominate the expansion of HII regions and cloud outflows. On its own, \cite{Reissl2018} find that the radiation pressure on dust grains almost never disrupts clouds with an \SFE below 50\% over a mass range from $10^4$ to $10^7$ \Msolar. \cite{Haworth2015} and \cite{Kim2018} also argue that photoionisation is indeed the dominant effect from photons in HII regions in both uniform density fields and turbulent clouds under the conditions studied in this paper.

All of our clouds are eventually destroyed and reach a plateau in \SFE before the first star dies. We thus do not expect supernovae to play a role in shaping these clouds. However, for longer-lived clouds such as in the \LMC \cite[e.g.][]{Kawamura2009}, some supernovae might occur within the lifetime of the cloud, affecting its evolution. In \cite{Geen2015a} we find that supernovae exploding into pre-existing HII regions transfer significantly more momentum to the \ISM, since a low-density region has been carved out. However, in \cite{Geen2016}, where we simulate a dense cloud and clumps remain embedded in the pre-supernova HII region, the supernova produces less momentum than in studies where gravoturbulence is less important, e.g. \cite{Iffrig2015}, \cite{Kim2015}, \cite{Martizzi2015} and \cite{Kortgen2016}.

All of our models use the Geneva rotating star evolution tables. A discussion of variable stellar rotation, binary interactions, etc, is beyond the scope of our model, but will, of course, be important if we wish to understand the full parameter space of star formation.

\section{Conclusions}
\label{conclusions}
In this paper we present the \textsc{yule} suite of simulations, which are 3D RMHD simulations of the same cloud performed with different initial mass function (IMF) sampling and initial turbulent velocity seeds, using an initial density distribution similar to clouds in the solar neighbourhood. We present a model for the expansion of HII regions into molecular clouds that shows good agreement with the behaviour of our clouds.

The main result of this paper is that the SFE of the cloud (measured as the total fraction of the initial mass converted to stars) can be anywhere between 6 and 23\% depending on the parameters that statistically vary within a cloud with a fixed set of initial global properties. This suggests that the SFE of clouds in the solar neighbourhood is difficult to predict. In addition to capturing a large parameter space of physical properties, simulations or other theoretical models must also capture the full range of statistical variation in the input parameters.

We compute a Bayesian model that identifies relationships between the SFE and various emergent properties of the cloud and cluster. When we vary the sampling of the IMF, the most significant parameter is the total number of photons emitted in the first freefall time, or approximately 2 Myr after the first star was formed. More photons give a lower SFE.

When varying the initial velocity structure, the most important parameter is the long axis of the ellipsoid fit to the cloud, since most of our clouds evolve more filamentary structures. Derived quantities such as the mean radius and volume were also significant. The more filamentary or extended the cloud, the lower the SFE. We suggest that the length of the cloud is important because the mass in our sample is fixed, so longer clouds are on average less compact. We find that sampling the whole cloud rather than only the densest regions provides a better relationship with the total SFE. This is in contrast to observational studies, which find a correlation between density peaks and recent star formation.

There is also an apparent link between the SFE and the distance massive stars travel. We suggest that this is because if feedback is inefficient (resulting in high SFE), dense gas clumps remain embedded for longer inside the cloud, and these clumps remain gravitationally bound to nearby stars. When these stars produce ionising radiation the surface of these clumps is evaporated, and the star-clump system is accelerated via the rocket effect, where momentum is conserved with the photoevaporated gas ejected in the opposite direction. This has the counterintuitive result that, in this regime, efficient feedback results in less cluster dispersal, while inefficient feedback causes the cluster to expand.

\section{Acknowlegements}
\label{acknowledgements}
\rev{We would like to thank the anonymous referee and the journal's editorial team for their work in improving the text. In addition,} the authors would like to thank Simon Glover, Thomas Greif, Sacha Hony, Ben Keller, Eric Pellegrini, Daniel Rahner and Stefan Reissl for their useful comments and discussions during the preparation of this work. 

This work was granted access to HPC resources of CINES under the allocation x2014047023 made by GENCI (Grand Equipement National de Calcul Intensif). The authors acknowledge support by the High Performance and Cloud Computing Group at the Zentrum f\"ur Datenverarbeitung of the University of T\"ubingen, the state of Baden-W\"urttemberg through bwHPC and the German Research Foundation (DFG) through grant no INST 37/935-1 FUGG. 

This work has been funded by the European Research Council under the European Community's Seventh Framework Programme (FP7/2007-2013). SG and RSK have received funding from Grant Agreement no. 339177 (STARLIGHT) of this programme. RSK further acknowledges support from the Deutsche Forschungsgemeinschaft in the Collaborative Research Centre SFB 881 ``The Milky Way System'' (subprojects B1, B2, and B8) and in the Priority Program SPP 1573 ``Physics of the Interstellar Medium'' (grant numbers KL 1358/18.1, KL 1358/19.2). PH has received funding from Grant Agreement no. 306483. JR acknowledges support from the ORAGE project from the Agence Nationale de la Recherche, grant ANR-14-CE33-0016-03.

% BIBLIOGRAPHY
 \bibliographystyle{mnras}
 \bibliography{samgeen}
 
 \appendix

 \section{List of stars sampled}
 \label{appendix:starlist}
 
 Here is a list of stellar masses formed in each of the simulations (see Table \ref{simtable}), in the order that they are formed (rounded to 0.1 \Msolar). Each of the values is given in \Msolar. In the \stars runs, where we vary the \IMF sampling, the simulations form the following masses
 
\textbf{STE} 39.9, 9.3, 9.0, 11.9, 22.7, 48.3

\textbf{GIL} 8.0, 18.7, 16.3, 25.6, 39.9, 11.1, 10.8

\textbf{STU} 21.7, 12.9, 9.8, 10.9, 14.5, 8.7, 13.0, 10.9, 8.3, 18.3, 25.8

\textbf{THV} 68.1, 12.3, 18.6, 18.0, 54.0

\textbf{POT} 39.6, 11.3, 22.3, 13.1, 18.8

\textbf{ASK} 30.8, 9.9, 10.8, 30.9, 10.2

\textbf{HUR} 31.2, 18.6, 16.4, 9.7, 14.6

\textbf{SKY} 8.8, 11.3, 13.4, 16.7, 13.8, 11.3, 9.4, 9.5, 55.0, 10.6, 21.0, 10.9

\textbf{BJU} 15.4, 8.3, 14.9, 89.8, 10.3

\textbf{GLU} 9.3, 8.7, 29.0, 107.2, 23.6

\textbf{GAT} 11.1, 9.9, 11.6, 12.7, 46.9, 8.4, 13.3, 12.4

\textbf{KET} 55.2, 8.1, 8.1, 9.9, 33.7, 12.3

\textbf{KER} 19.4, 14.1, 26.3, 8.5, 9.3, 10.9, 8.5, 17.8

In the \turb runs, we use the same base list of stellar masses, although different simulations have different \SFE, resulting in longer or shorter lists (note that in our star formation recipe, we form one massive star every time 120 \Msolar is accreted onto all sink particles):

 \textbf{GRY} 30.8, 9.9, 10.8, 30.9, 10.2, 13.7, 34.4, 58.3, 9.7, 16.5

\textbf{JOL} 30.8, 9.9, 10.8, 30.9, 10.2, 13.7, 34.4

\textbf{JOU} 30.8, 9.9, 10.8, 30.9, 10.2, 13.7, 34.4, 58.3, 9.7

\textbf{GAL} 30.8, 9.9, 10.8, 30.9, 10.2, 13.7, 34.4

\textbf{TIO} 30.8, 9.9, 10.8, 30.9, 10.2, 13.7, 34.4, 58.3, 9.7, 16.5, 43.0, 18.6

\textbf{CAG} 30.8, 9.9, 10.8, 30.9, 10.2, 13.7

\textbf{SNE} 30.8, 9.9, 10.8, 30.9, 10.2, 13.7, 34.4, 58.3, 9.7

\textbf{BEF} 30.8, 9.9, 10.8, 30.9, 10.2, 13.7, 34.4, 58.3, 9.7, 16.5, 43.0, 18.6, 60.8, 8.7, 9.8, 83.5, 9.2, 16.0

\textbf{STL} 30.8, 9.9, 10.8, 30.9, 10.2, 13.7, 34.4, 58.3, 9.7, 16.5, 43.0

\textbf{MAR} 30.8, 9.9, 10.8, 30.9, 10.2, 13.7, 34.4, 58.3, 9.7, 16.5, 43.0

\textbf{TAN} 30.8, 9.9, 10.8, 30.9, 10.2, 13.7, 34.4, 58.3

\textbf{OLD} 30.8, 9.9, 10.8, 30.9, 10.2

\textbf{YAG} 30.8, 9.9, 10.8, 30.9

\section{Stellar Evolution Model}
\label{appendix:starmodel}

\begin{figure}
	\centerline{\includegraphics[width=0.9\hsize]{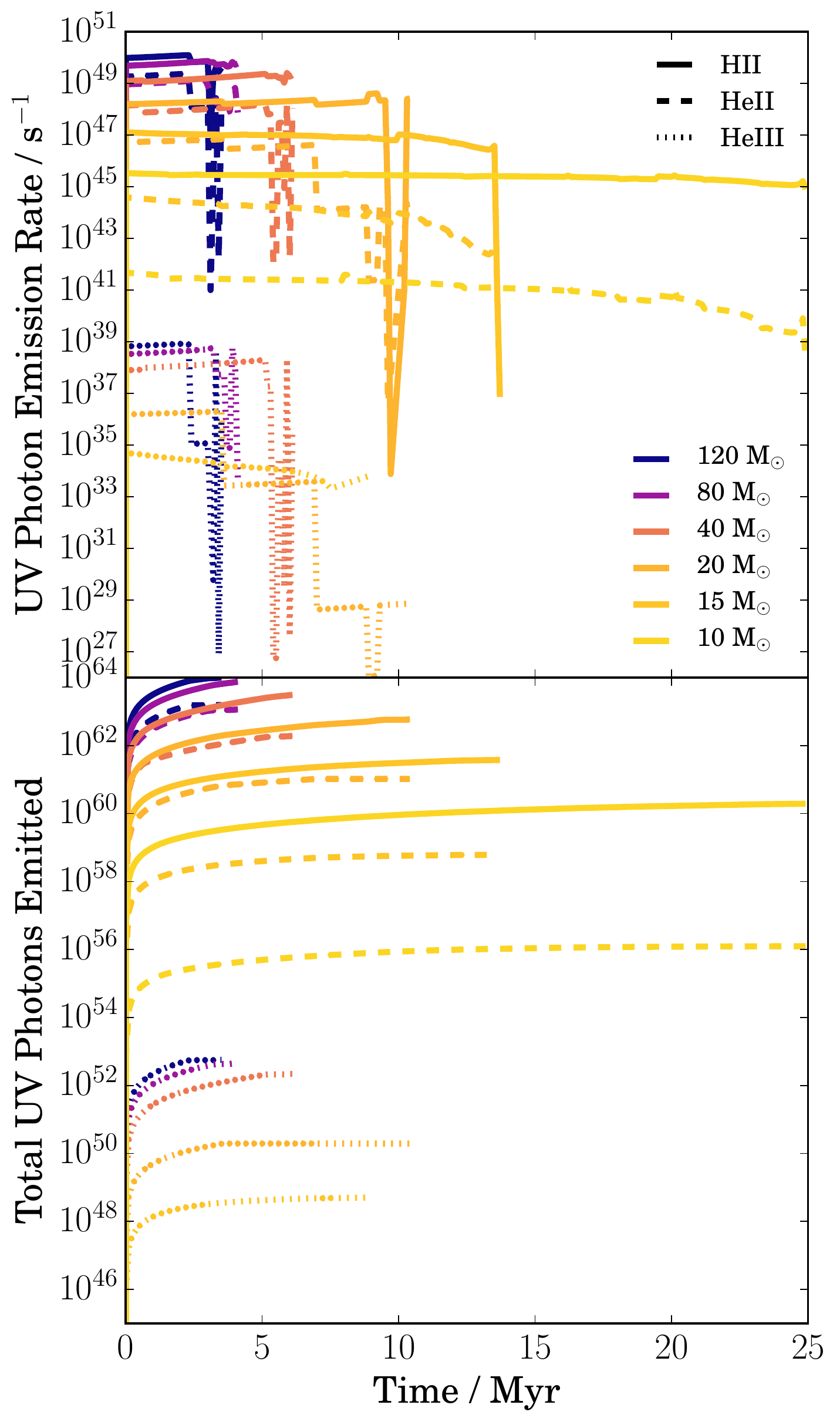}}
	\caption{Ionising UV photon emission rates versus time for a sample of star masses in stellar evolution model (our model interpolates between tracks sampled every 5 \protect\Msolar from 5 to 120 \protect\Msolar). Solid lines show the lowest energy bins that can ionise hydrogen, dashed lines show the next highest energy bin that can ionise helium once, and dotted lines show the highest energy bin that can ionise helium twice.}
	\label{stellarevolution}
\end{figure}

In Figure \ref{stellarevolution} we plot the total ionising UV photon emission rates as a function of time for a selection of massive stars as modelled in this paper. Details of the model are given in Section \ref{simulations:stellarevolution}.

\section{Ionisation Front Expansion}
\label{appendix:ionfront}

In this Appendix we provide a brief overview of the expansion of photoionised HII regions in small ($10^4$ \Msolar) molecular clouds, in order to provide a theoretical context for the rest of the paper. For larger or longer-lived clouds, processes like stellar winds and supernovae become important \citep{Rahner2017}.

\subsection{Modes of Expansion of Photoionised HII Regions}

The expansion of ionisation fronts occurs in two modes, called R-type and D-type \citep{KahnF.D.1954}. In the R-type mode, the expansion is more-or-less hydrostatic, and traces the absorption of ionising photons emitted from the star by the surrounding gas, tending towards an equilibrium at $r_s$, or the Str\"omgren radius, where all of the ionising photons are absorbed by the ambient medium.

The R-type solution for a radius $r_i$ in a uniform medium at time $t$ can be solved as $r_i(t) = r_{s} \left (1 - e^{-n_0 \alpha_B t} \right )$ where $n_0$ is the hydrogen number density of the external medium and $\alpha_B$ is the recombination rate (here, $3\times10^{-13}$ cm$^3$/s). The recombination time $t_{rec} = 1/(n_0 \alpha_B)$.

\cite{Shapiro2006} derive more general forms for this equation in different environments, as well as for non-infinite light speeds (which is more relevant in cosmological contexts). Regardless of the exact solution, the evolution of this phase is typically much shorter than the freefall time of the cloud and the lifetime of the massive stars. For $n_0=100$ cm$^{-3}$, $t_{rec}\simeq1000$ years, while the freefall time of our cloud is 4.2 Myr, comparable to the lifetime of the most massive stars.

In the D-type mode, the ionisation front has reached photoionisation equilibrium, and a hydrodynamic expansion is driven by the pressure difference between the photoheated gas (which has a temperature of approximately $10^4$ K) and the external medium, while maintaining ionisation equilibrium. We discuss this expansion in some detail in \cite{Geen2015b}, while in \cite{Geen2016} we note that in a self-gravitating, turbulent cloud, this expansion occurs on the scale of the freefall time in the ambient medium.

\subsection{Expansion in a Power Law Density Field}

In Figure \ref{rayprofs} we plot the density structure around the first star in the \stars simulations, and around the geometric centre of the cloud. While the cloud as a whole can, as in \cite{Geen2016}, be treated as having a mostly flat density structure, the first star ``sees'' a power law density profile $n(r) \propto r^{-w}$ (where $w=1.71\pm0.05$ here when producing a power law fit to the mean density in Figure \ref{rayprofs}).

In \cite{Geen2015b}, we solve the expansion of a D-type HII region in such a density field to give
\begin{equation}
r_i \propto t^{\psi} S_*^{\psi/4}
\label{powerlawradius}
\end{equation}
where $\psi \equiv 4 / (7 - 2 w)$. 

The momentum of the shell can also be calculated by assuming that all of the mass swept up by the ionisation front is contained in the shell and calculating $M(<r_i) \dot{r_i}$. For a uniform cloud where $w=0$, these solutions reduce to the equations given in \cite{Matzner2002}, which are simplifications of \cite{SpitzerLyman1978} and \cite{Dyson1980}.

One important aspect is raised by \cite{Franco1990} \citep[see also][]{Shu2002}, who argue that for $w>1.5$, the HII region enters a ``champagne'' flow phase \citep{TenorioTagle1979,Whitworth1979}, in which the ionisation front rapidly bursts out of the cloud. This happens when the ionisation front leaves ionisation equilibrium. Ionisation equilibrium in this power law density field is found by balancing the number of photons emitted by the star per unit time with the number of recombinations in the density field around the star
\begin{equation}
S_* = \frac{4}{3} \pi r_i^3 n_i^2 \alpha_B 
\label{stromgrenradius}
\end{equation} where $n_i$ is the density in the ionised gas. The gas can remain in the D-type phase as long $n_i$ does not exceed the integrated density in the unperturbed gas $n(r)$ up to $r_i$ at which point all of the mass in the HII region including the shell is photoionised and the ionisation front re-enters the R-type phase. Solving Equation 4 in \cite{Alvarez2006} gives $r_s^{3-2w} \propto S_*$. As long as $w < 3/2$, this condition is never met. Otherwise, a ``breakout'' can occur. \cite{Alvarez2006} give a breakout radius $r_B$ and time $t_B$ at which this occurs. Using equation 6 in their paper (assuming our cloud can be approximated as an isothermal profile), we calculate $r_B\sim10$ pc for our strongest source with $S_*\simeq10^{49}$ s$^{-1}$, which is roughly the size of our cloud (see Figure \ref{rayprofs}). We thus estimate that our ionisation front should, in principle, not leave the D-type phase before it escapes the cloud. We compare these models to our simulations in Section \ref{globalprops:radius}.

\section {Additional Statistical Model Tables}
\label{appendix:morestats}

Here we give the full model tables described in Section \ref{correlations:models}. See Tables \ref{tablestats_turb1}, \ref{tablestats_turb2}, \ref{tablestats_turb3} and \ref{tablestats_stars}.

\begin{table*}
	\centering
	\begin{tabular}{lllllll}
		
		\thead{Model \\ structure} & \thead{$\mathbf{t/t_{ff}}$} & \thead{Fixed \\ effect} & \thead{Posterior \\ mean} & \thead{Lower \\ 95\% CI} & \thead{Upper \\ 95\% CI} & \thead{DIC} \\
		\hline
		Full  & 0.5 & Intercept    & 4.053          & 2.060         & 6.075         & -27.4                       \\
		&     & $V_E$            & -1.115         & -1.850        & -0.371        &                             \\
		Null  & 0.5 & Intercept    & 1.029          & 0.931         & 1.126         & -22.4                       \\
		\hline
		Full  & 1.0 & Intercept    & 3.602          & 1.150         & 6.048         & -24.6                       \\
		&     & $V_E$            & -0.868         & -1.683        & -0.032        &                             \\
		Null  & 1.0 & Intercept    & 1.029          & 0.931         & 1.127         & -21.9                       \\
		\hline
		Full  & 1.5 & Intercept    & 0.951          & 0.015         & 1.859         & -20.4                       \\
		&     & $V_E$            & 0.025          & 0.277         & 0.314         &                             \\
		Null  & 1.5 & Intercept    & 1.029          & 0.932         & 1.130         & -22.4                       \\
		\hline
		Full  & 2.0 & Intercept    & 1.252          & 0.685         & 1.864         & -21.4                       \\
		&     & $V_E$            & 0.071          & -0.268        & 0.102         &                             \\
		Null  & 2.0 & Intercept    & 1.029          & 0.929         & 1.129         & -21.8                      
	\end{tabular}
	\caption{Table of model results for \turb parameter $V_E$ at 0.5, 1, 1.5 and 2.0 $t_{ff}$, as in Table \ref{tableX}.}
	\label{tablestats_turb1}
\end{table*}

\begin{table*}
	\centering
	\begin{tabular}{lllllll}
		
		\thead{Model \\ structure} & \thead{$\mathbf{t/t_{ff}}$} & \thead{Fixed \\ effect} & \thead{Posterior \\ mean} & \thead{Lower \\ 95\% CI} & \thead{Upper \\ 95\% CI} & \thead{DIC} \\
		\hline
		Full  & 0.5 & Intercept    & 3.62           & 2.05          & 5.23          & -28.0                       \\
		&     & $R_E$            & -2.68          & -4.37         & -1.06         &                             \\
		Null  & 0.5 & Intercept    & 1.03           & 0.93          & 1.13          & -22.4                       \\
		\hline
		Full  & 1.0 & Intercept    & 3.247          & 1.915         & 4.546         & -28.5                       \\
		&     & $R_E$            & -2.015         & -3.155        & -0.768        &                             \\
		Null  & 1.0 & Intercept    & 1.03           & 0.93          & 1.13          & -21.9                       \\
		\hline
		Full  & 1.5 & Intercept    & 1.58           & 0.54          & 2.62          & -22.1                       \\
		&     & $R_E$            & -0.45          & -1.276        & 0.42          &                             \\
		Null  & 1.5 & Intercept    & 1.03           & 0.93          & 1.13          & -22.4                       \\
		\hline
		Full  & 2.0 & Intercept    & 1.01           & 0.27          & 1.78          & -21.0                       \\
		&     & $R_E$            & 0.01           & -0.52         & 0.58          &                             \\
		Null  & 2.0 & Intercept    & 1.03           & 0.93          & 1.13          & -21.8  
	\end{tabular}
	\caption{Table of model results for \turb parameter $R_E$ at 0.5, 1, 1.5 and 2.0 $t_{ff}$, as in Table \ref{tableX}.}
	\label{tablestats_turb2}
\end{table*}

\begin{table*}
	\centering
	\begin{tabular}{lllllll}
		\thead{Model \\ structure} & \thead{$\mathbf{t/t_{ff}}$} & \thead{Fixed \\ effect} & \thead{Posterior \\ mean} & \thead{Lower \\ 95\% CI} & \thead{Upper \\ 95\% CI} & \thead{DIC} \\
		\hline
		Full  & 0.5 & Intercept    & 2.036          & -1.221        & 5.227         & -25.8                       \\
		&     & $T$            & -0.906         & -3.843        & 2.061         &                             \\
		&     & $L_E$            & -0.552         & -4.504        & 3.183         &                             \\
		&     & $M_E$            & -1.556         & -7.052        & 3.571         &                             \\
		&     & $S_E$            & 0.602          & -3.844        & 5.557         &                             \\
		Null  & 0.5 & Intercept    & 1.029          & 0.931         & 1.126         & -22.4                       \\
		Best  & 0.5 & Intercept    & 3.621          & 2.046         & 5.232         & -28.4                       \\
		&     & $L_E$            & -2.683         & -4.367        & -1.056        &                             \\
		\hline
		Full  & 1.0 & Intercept    & 2.983          & 1.037         & 4.825         & -25.5                       \\
		&     & $T$            & -0.398         & -2.180        & 1.471         &                             \\
		&     & $L_E$            & -1.104         & -3.350        & 0.885         &                             \\
		&     & $M_E$            & -1.073         & -4.577        & 2.662         &                             \\
		&     & $S_E$            & -0.126         & -1.583        & 1.338         &                             \\
		Null  & 1.0 & Intercept    & 1.029          & 0.931         & 1.127         & -21.9                       \\
		Best  & 1.0 & Intercept    & 3.247          & 1.915         & 4.546         & -29.2                       \\
		&     & $L_E$            & -2.015         & -3.155        & -0.768        &                             \\
		\hline
		Full  & 1.5 & Intercept    & 1.478          & 0.590         & 2.348         & -23.4                       \\
		&     & $T$            & -1.410         & -4.098        & 1.675         &                             \\
		&     & $L_E$            & 0.149          & -2.142        & 2.357         &                             \\
		&     & $M_E$            & -1.136         & -4.600        & 2.005         &                             \\
		&     & $S_E$            & 0.400          & -0.229        & 1.002         &                             \\
		Null  & 1.5 & Intercept    & 1.029          & 0.932         & 1.130         & -22.4                       \\
		Best  & 1.5 & Intercept    & 0.906          & 0.714         & 1.119         & -22.5                       \\
		&     & $T$            & -0.704         & -1.717        & 0.285         &                             \\
		\hline
		Full  & 2.0 & Intercept    & 1.060          & -0.068        & 2.122         & -19.8                       \\
		&     & $T$            & 0.138          & -1.453        & 1.884         &                             \\
		&     & $L_E$            & 0.166          & -0.771        & 1.177         &                             \\
		&     & $M_E$            & -0.239         & -0.982        & 0.522         &                             \\
		&     & $S_E$            & -0.076         & -0.888        & 0.768         &                             \\
		Null  & 2.0 & Intercept    & 1.029          & 0.929         & 1.129         & -21.8                       \\
		Best  & 2.0 & Intercept    & 1.237          & 0.881         & 1.556         & -22.0                       \\
		&     & $M_E$            & -0.235         & -0.608        & 0.129         &  
	\end{tabular}
	\caption{Table of model results for \turb parameters $T$, $L_E$, $M_E$ and $S_E$ at 0.5, 1, 1.5 and 2.0 $t_{ff}$, as in Table \ref{tableX}. DIC is defined in Section \ref{correlations:models}.}
	\label{tablestats_turb3}
\end{table*}

\begin{table*}
	\centering
	\begin{tabular}{lllllll}
		
		\thead{Model \\ structure} & \thead{Fixed \\ effect} & \thead{Posterior \\ mean} & \thead{Lower \\ 95\% CI} & \thead{Upper \\ 95\% CI} & \thead{DIC} \\
		\hline
		Full & Intercept   & 8.01148  & 2.89428 & 13.6   & -34.103 \\
		& $M_{max}$  & 0.07901  & -0.349  & 0.5    &         \\
		& $R_{RMS}$ & 0.19739  & -0.821  & 1.312  &         \\
		& $M_{first}$   & -0.13375 & -0.382  & 0.110  &         \\
		& $N_{ff}$ & -0.11745 & -0.215  & -0.020 &         \\
		Null & Intercept   & 0.9169   & 0.838   & 0.993  & -26.1   \\
		Best & Intercept   & 8.23102  & 4.438   & 12.093 & -35.2   \\
		& $N_{ff}$ & -0.11925 & -0.182  & -0.057 &  
	\end{tabular}
	\caption{Table of model results for \stars parameters, as in Table \ref{tableX}. DIC is defined in Section \ref{correlations:models}.}
	\label{tablestats_stars}
\end{table*}

\bsp	% typesetting comment
\label{lastpage}
\end{document}